# Colloquium: Topological Band Theory

A. Bansil[1]*, Hsin Lin[1,2,3], Tanmoy Das[2,3,4]

*Corresponding author: A. Bansil (bansil@neu.edu)
[1]Department of Physics, Northeastern University, Boston, MA 02115
[2]Centre for Advanced 2D Materials and Graphene Research Centre, National University of Singapore, Singapore 117546
[3]Department of Physics, National University of Singapore, Singapore 117542
[4]Theoretical Division, Los Alamos National Laboratory, Los Alamos, New Mexico 87545

The first-principles band theory paradigm has been a key player not only in the process of discovering new classes of topologically interesting materials, but also for identifying salient characteristics of topological states, enabling direct and sharpened confrontation between theory and experiment. We begin this review by discussing underpinnings of the topological band theory, which basically involves a layer of analysis and interpretation for assessing topological properties of band structures beyond the standard band theory construct. Methods for evaluating topological invariants are delineated, including crystals without inversion symmetry and interacting systems. The extent to which theoretically predicted properties and protections of topological states have been verified experimentally is discussed, including work on topological crystalline insulators, disorder/interaction driven topological insulators (TIs), topological superconductors, Weyl semimetal phases, and topological phase transitions. Successful strategies for new materials discovery process are outlined. A comprehensive survey of currently predicted 2D and 3D topological materials is provided. This includes binary, ternary and quaternary compounds, transition metal and $f$-electron materials, Weyl and 3D Dirac semimetals, complex oxides, organometallics, skutterudites and antiperovskites. Also included is the emerging area of 2D atomically thin films beyond graphene of various elements and their alloys, functional thin films, multilayer systems, and ultra-thin films of 3D TIs, all of which hold exciting promise of wide-ranging applications. We conclude by giving a perspective on research directions where further work will broadly benefit the topological materials field.

## CONTENTS

## I. INTRODUCTION

Successful prediction of new TI materials is perhaps the most spectacular triumph of the modern first-principles band theory.(Hohenberg and Kohn 1964; Kohn and Sham 1965) TIs are an exotic state of quantum matter, which is distinct from the ordinary insulators in that even though electrons cannot conduct in the bulk of the material, surfaces of three-dimensional (3D) TIs and edges of two-dimensional (2D) TIs support metallic or conducting electronic states protected by constraints of time-reversal symmetry (TRS) (Kane and Mele 2005a; Kane and Mele 2005b; Bernevig, Hughes, and Zhang 2006; Fu, Kane, and Mele 2007; Moore and Balents 2007; Fu and Kane 2007; Roy 2009). The gapless topological surface/edge states are forged when bonding forces involved in crystal formation are modified in a special manner through relativistic spin-orbit interactions. These metallic states are unique in that they exhibit chirality or locking of the directions of spin and momentum, and are not allowed to scatter backwards (no 'U-turns') in the absence of magnetic impurities. The experimental discovery of TIs (Konig et al. 2007; Hsieh et al. 2008; Xia et al. 2009; Hsieh, Xia, Qian, Wray, Dil, et al. 2009; Chen et al. 2009; Hsieh, Xia, Qian, Wray, Meier, et al. 2009) has focused attention on wide-ranging phenomena in materials driven by spin-orbit coupling (SOC) effects well beyond the traditional role of SOC in determining magnetic properties. In this sense topological materials represent the emergence of spin in quantum matter from being a 'spectator' to a 'driver', and a shift in focus from Schrödinger equation to Dirac equation as the controlling physics. TIs have brought topological considerations front and center in discussing the physics of materials, building on the earlier work on topology and topological orders in condensed matter systems (Thouless et al. 1982; Haldane 1988; Volovik 1988; Wen 1995; Zhang and Hu 2001; Murakami, Nagaosa, and Zhang 2004; Sinova et al. 2004). The special symmetry protected electronic states in the TIs hold the exciting promise of providing revolutionary new platforms for exploring fundamental science questions, including novel spin-textures and exotic superconducting states as well as for the realization of multifunctional topological devices for thermoelectric, spintronics, information processing and other applications. (Hasan and Kane 2010; Qi and Zhang 2011; Mas-Ballesté et al. 2011; Butler et al. 2013; Tsai et al. 2013)

By examining how band structures evolve under spin-orbit interaction, many topologically interesting materials have been predicted. Theoretically predicted 3D TIs span binary, ternary and quaternary compounds, transition metal and $f$-electron systems, Weyl and 3D Dirac semimetals, complex oxides, organometallics, skutterudites and antiperovskites as summarized in the materials inventory given in the Appendix. In many cases, even though the pristine phase is found to be topologically trivial, computations show that the material could be coaxed into assuming a non-trivial phase through alloying, strain or confinement. Among the practically realized materials, nontrivial compounds were first predicted theoretically before experimental verification, including $Bi_{1-x}Sb_x$ (Fu and Kane 2007; Hsieh et al. 2008) for $Z_2$ phase (Sec. IV.B1), SnTe (Hsieh et al. 2012; Tanaka et al. 2012; Dziawa et al. 2012; Xu, Liu, et al. 2012) as a topological crystalline insulator (Sec. IV.B6), $Cd_3As_2$ and $Na_3Bi$ (Wang, Sun, et al. 2012; Wang, Weng, et al. 2013; Liu, Zhou, et al. 2014; Borisenko et al. 2014; Neupane, Xu, et al. 2014; Xu, Liu, et al. 2015; Ali et al. 2014) as 3D Dirac cone semimetals, and TaAs (Huang et al. 2015; Weng et al. 2015; Xu, Belopolski, et al. 2015; Lv et al. 2015; Xu, Alidoust, et al. 2015; Zhang et al. 2015) as a Weyl semimetal (Sec. H). In the widely studied $Bi_2Se_3$ (Xia et al. 2009; Zhang, Liu, et al. 2009) and $GeBiTe_4$ (Xu et al. 2010) TIs, experimental verification and theoretical prediction occurred simultaneously. Many theoretically predicted TIs however have not been realized experimentally. The early experimental work on the first generation elemental TIs, Bi/Sb, quickly gave way to the second generation binary materials, $Bi_2Se_3$, $Bi_2Te_3$, $Sb_2Te_3$ and their alloys, followed by work on ternary and quaternary materials, which offer greater flexibility and tunability with respect to lattice parameters, chemical compositions, band gaps and transport properties.

Despite the progress made in synthesizing 3D TIs, the materials realization of 2D TIs or quantum spin Hall (QSH) insulators is still limited to the HgTe/CdTe (König et al. 2007)and InAs/GaSb/AlSb (Knez, Du, and Sullivan 2011) quantum well systems with small band gaps of ~4-10 meV. In both these 2D materials, theoretical prediction preceded experimental verification. (Bernevig, Hughes, and Zhang 2006; Liu et al. 2008) On the other hand, first principles computations on atomically thin films of many elements and their alloys, and ultra-thin films of most 3D TIs, yield numerous stable structures capable of supporting the QSH phase with band gaps large enough in many cases for room temperature applications, see the Appendix for an inventory. In sharp contrast to the case of graphene (Castro Neto et al. 2009; Das Sarma et al. 2011), which possesses a flat structure in its pristine form, the structure of most atomically thin films is naturally buckled so that their inversion symmetry can be broken by an external electric field.(Mas-Ballesté et al. 2011; Butler et al. 2013) A freestanding silicene sheet (one atom thick crystal of Si), for example, can harbor a rich phase diagram as a function of external electric and magnetic fields in which it transitions from being a band insulator, to a quantum anomalous or spin Hall insulator, to a valley polarized metal, to a spin valley polarized metallic phase. (Tsai et al. 2013; Ezawa 2012; Liu, Feng, and Yao 2011; Drummond, Zólyomi, and Fal'ko 2012; Tabert and Nicol 2013) A silicene nanoribbon could thus be used to manipulate spin-polarized currents (Tsai et al. 2013; Gupta et al. 2014; Saari et al. 2014) via gating without the need to switch magnetic fields. Moreover, a rich tapestry of morphologies and topological characteristics is generated when we consider layer-by-layer growth of thin films on various substrates(Mas-Ballesté et al. 2011; Butler et al. 2013; Huang, Chuang, et al. 2013).

While a great deal of the existing literature on TIs has been concentrated on the $Z_2$ insulators in which the gapless surface/edge states are protected by the TRS, interest in exploring the role of crystal symmetries in creating protected states more generally has been growing. This has given birth to a new class of TIs called topological crystalline insulators (TCIs). (Fu 2011) In this connection, mirror symmetry has received special attention, on the basis of which band theory computations predicted SnTe and $Pb_{1-x}Sn_x(Se,Te)$ alloys to harbor a TCI state, (Hsieh et al. 2012) which was verified essentially immediately afterwards by three different experimental groups. (Dziawa et al. 2012; Tanaka et al. 2012; Xu, Liu, et al. 2012) The Dirac cone structure and the associated spin-texture of surface states in a TCI is quite distinct from that in the more common TIs. (Fu 2011; Wang, Tsai, et al. 2013) For example, the TCI SnTe supports an even number (not an odd number that is hallmark of a TI) of metallic Dirac cone states on crystal surfaces, which are symmetric under reflections in the {110} planes. $Pb_{1-x}Sn_x(Se,Te)$ system is the first and still the only materials realization of a TCI, see the Appendix for other materials predictions.

Topological materials practically realized to date are essentially 'weakly correlated' in the sense that the standard density functional theory (DFT) based independent particle picture (Hohenberg and Kohn 1964) (Kohn and Sham 1965) provides a reasonably robust description of their electronic structures, limitations of the DFT in correctly capturing the size of the band gap in semiconductors and insulators notwithstanding. Methods for treating strong electron correlation effects under the rubrics of LDA+U, GW, LDA+DMFT, and various type of exchange-correlation functionals have been reviewed extensively in the literature. (Peverati and Truhlar 2014; Capelle and Campo Jr. 2013; Held et al. 2006;

Maier et al. 2005; Kotliar et al. 2006; Das, Markiewicz, and Bansil 2014) It is natural to expect that the combined effects of strong correlations and SOC would give rise to entirely new classes of TIs. Here, Mott (Zhang, Zhang, Wang, et al. 2012; Deng, Haule, and Kotliar 2013) and Kondo insulators (Dzero et al. 2010; Weng, Zhao, et al. 2014) provide a natural breeding ground for finding correlated TIs. The iridates, which exhibit many exotic phenomena through the interplay of $5d$ electrons and strong SOC are drawing interest as candidate materials, although their topological nature remains elusive.(Wan et al. 2011; Yang, Lu, and Ran 2011; Carter et al. 2012) Among the $f$-electron systems, attention has been focused on $SmB_6$,(Cooley et al. 1995; Frantzeskakis et al. 2013; Jiang et al. 2013; Kim, Xia, and Fisk 2014; Xu, Shi, et al. 2013; Neupane, Alidoust, et al. 2013) which might support a topological Kondo insulator state as the bulk becomes insulating at low temperatures.

Many aspects of the electronic structures and properties of topological materials are difficult or impractical to model within a totally first-principles framework, and as a result, a variety of effective model Hamiltonians are invoked frequently in the field. Material specificity can be obtained by choosing parameters entering the model Hamiltonian to mimic appropriate first-principles results to varying degrees, although generic features can often be captured via minimal models consistent with the symmetries inherent in particular problems. In this way, new insights have been enabled in the characteristics of topological superconductors (Fu and Kane 2008; Qi, Hughes, et al. 2009) and their interfaces with magnetic and non-magnetic materials, (Qi, Li, et al. 2009; Wray et al. 2011; Oroszlány and Cortijo 2012; Wei et al. 2013) effects of external electric and magnetic fields on 2D and 3D TIs,(Cho et al. 2011; Zhu, Richter, et al. 2013; Essin, Moore, and Vanderbilt 2009; Ojanen 2012; Baasanjav, Tretiakov, and Nomura 2013) evolution of electronic states with dimensionality, (Bansal et al. 2012; Glinka et al. 2013; Wang, Liu, et al. 2012; Kim, Brahlek, et al. 2011; Vargas et al. 2014) and various exotic quantum phenomena possible in the TIs (e.g., Majorana fermions (Fu and Kane 2008; Elliott and Franz 2015; Roy and Kallin 2008; Kitaev 2009; Schnyder et al. 2008), axions,(Essin, Moore, and Vanderbilt 2009; Li et al. 2010) magnetic monopoles, (Qi, Li, et al. 2009) fractional excitations (Wen 1995; Grushin et al. 2012; Li, Liu, et al. 2014; Teo and Kane 2014), which are not amenable to treatment on a first-principles basis.

This review is organized as follows. Sec. II discusses underpinnings of the topological band theory. We delineate methods for characterizing the topology of the bulk band structure in terms of the $Z_2$ invariants, including surface/edge state computations and the treatment of interacting systems (Secs. IIA-D). Role of model Hamiltonians in addressing various classes of problems, which are not amenable to first-principles treatment, is pointed out (Sec. IIE), and the extent to which theoretically predicted properties and protections of topological states have been verified experimentally is discussed (Sec. IIF). Sec. III turns to topological phases/transitions in quantum matter more broadly, and considers: topological crystalline insulators (Sec. IIIA); disorder/interaction driven TIs (Sec. IIIB); topological superconductors (Sec. IIIC); Weyl and 3D Dirac semimetal phases (Sec. IIID); and, topological phase transitions (Sec. IIIE). Sec. IV provides a survey of currently predicted topological materials. Successful strategies for new materials discovery process are summarized (Sec. IVA), and salient features of different types of 3D and 2D topological materials are highlighted, including 2D thin-film materials (Sec. IVD). The Appendix provides an inventory of 2D and 3D topological materials reported in the literature, along with pertinent references in each case. Although this review is focused on band theory work, we have made an effort to provide a broader perspective to the extent possible within space limitations.

Topological insulators have been reviewed previously in the *Reviews of Modern Physics* (Hasan and Kane 2010; Qi and Zhang 2011). A number of other reviews on various aspects of TIs have appeared elsewhere. (Bardarson and Moore 2013; Beenakker 2013; Dzero and Galitski 2013; Fruchart and Carpentier 2013; Ando 2013; Zhang and Zhang 2013; Yan and Zhang 2012; Feng and Yao 2012; Okuda and Kimura 2013) The literature cited in the present review should not be considered exhaustive, although it is fairly complete as of the submission date, and includes some subsequent updating. We have tended to cite more recent publications in many cases as entry points for accessing earlier literature.

## II. Underpinnings of Topological Band Theory

Electronic spectrum of a crystal can be organized in the form of energy bands as a function of the crystal momentum $\mathbf{k}$, which is guaranteed to be a good quantum number due to the translational symmetry of the lattice. Practical band structure computations invoke a one-particle or an effective independent electron picture within the framework of the density functional theory (DFT) proposed by Hohenberg and Kohn, (Hohenberg and Kohn 1964) and its implementation by Kohn and Sham. (Kohn and Sham 1965) The DFT based band structure construct has provided a remarkable ordering principle for understanding wide-ranging properties of metals, semiconductors and insulators, and an encoding of the materials 'genome'. In a metal, electrons occupy partially filled bands, which stretch across the Fermi energy, in an insulator filled and empty states are separated by a large band gap, while in a semiconductor this band gap is small enough to be bridged by thermal or other excitations. A band insulator, which is described by a gap in the one-particle band structure, is not the only type of insulator, and many other possibilities arise when additional broken or invariant symmetries are taken into account.

Lattice symmetries have always played a substantial role in the band theory of solids. Topological band theory, especially insofar as first principles band structures are concerned, expands the consideration of symmetries to encompass the TRS. In the context of tight-binding model Hamiltonians, however, a further expansion is possible by including effects of the particle-hole symmetry characteristic of the superconducting state. Just as lattice symmetries have led to classification schemes based on point and space groups of the lattice, inclusion of time-reversal and/or particle-hole symmetries in the mix yields new schemes for classifying allowed exotic states of quantum matter. (Hasan and Kane 2010; Qi and Zhang 2011; Kitaev 2009; Schnyder et al. 2008; Fang, Gilbert, and Bernevig 2012; Slager et al. 2013; Chiu, Yao, and Ryu 2013; Hughes, Prodan, and Bernevig 2011; Shiozaki and Sato 2014)

Topological band theory in effect involves a layer of analysis and interpretation for assessing the topological characteristics of the bulk electronic spectrum, which sits on top of a standard band structure calculation. The spin-orbit interaction must be included in the computation. Non-relativistic or semi-relativistic treatment, which cannot mix up- and down-spin states, is generally not sufficient. Even when the existence of topological states can be inferred from symmetry considerations, appropriate surface/edge state computations must be undertaken to determine the number, spin-texture, and location in energy and momentum of these states on specific surfaces/edges. Theoretical predictions along these lines are playing a key role in identifying particular topological states experimentally, and in the discovery of new topological materials.

### IIA. Characterizing a TI within the DFT

The telltale signature of the possibility that the band structure of an insulator might harbor a non-trivial topological phase is the inversion of energy levels with respect to their natural order around the bandgap at high symmetry points in the Brillouin zone (BZ) as shown in the schematic drawing of Fig. 1. Several points should be understood clearly here: (1) Either the band inversion or the opening of the band gap must be driven by the SOC; (2) 'Natural order' means the order of appropriate atomic levels. These are s- and p- orbitals in many semiconductors where the energy of the s-level lies above that of p-levels, but the s-orbital experiences an attractive relativistic potential strong enough to pull it below the p-orbitals in the solid. The inversion can however involve any pair of orbitals. Examples of materials, which involve orbitals with higher principal quantum numbers or with different magnetic quantum numbers in the inversion process will be seen in Sec. IV; (3) Finally, an inverted band structure by itself does not prove the existence of a non-trivial phase, but further analysis is required to establish its topological nature.

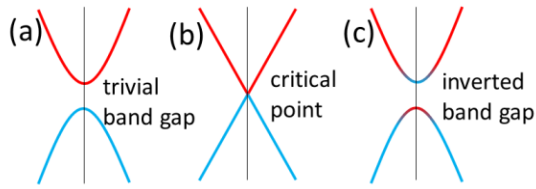

Fig. 1. Schematic band inversion between two bands derived from the natural order of atomic levels. The trivial band gap in (a) closes at a critical point in (b), and reopens inverted in (c) with the two states swapping their orbital characters at the symmetry point.

Three different types of approaches, which are discussed at length in the following sections, can be used to assess the topological characteristics of the band structure: (1) Compute $Z_2$ topological invariants, which encode time-reversal invariance properties of the bulk band structure; (2) Use adiabatic continuity arguments where one attempts to connect the unknown band structure to a known topological or non-topological band structure via a series of adiabatic or slowly varying transformations of the Hamiltonian without closing the band gap; (3) Directly compute the spectrum of surface/edge states which connect the bulk conduction and valence bands.

## IIB. Computation of $Z_2$ Topological Invariants

Among the various formulations for the computation of $Z_2$ topological invariants (Kane and Mele 2005a) Fu-Kane criterion (Fu, Kane, and Mele 2007) is especially well-suited for analyzing band structures of crystals with inversion symmetry. Evaluation of $Z_2$ in systems without inversion symmetry and interacting systems is discussed below.

Fu-Kane approach connects the $Z_2$ invariants to the matrix elements of Bloch wave functions at time-reversal invariant momentum (TRIM) points in the BZ. There are four TRIM points in the 2D BZ and eight in 3D. TRS yields one unique $Z_2$ invariant, $\nu$, in 2D, but four distinct $Z_2$ invariants ($\nu_0$;$\nu_1\nu_2\nu_3$) in 3D. In 2D, the two values of $\nu$ separate topological (or QSH) and non-topological states. The situation in 3D is more complicated with the involvement of four $Z_2$ invariants. Here, $\nu_0 = 1$ identifies a 'strong' TI in the sense that the system is robust against weak time reversal invariant perturbations, and any of its surfaces is guaranteed to host gapless surface bands. An ordinary insulator is obtained when all four invariants are zero. In the mixed case where $\nu_0 = 0$ and at least one of the indices $\nu_1$, $\nu_2$, or $\nu_3$ is non-zero, the 3D material can be viewed as a stacking of 2D TIs, and it is considered a 'weak' TI in the sense that it is less robust against perturbations.

Formally, Fu and Kane introduce the quantities $\delta_i = \text{Pf}[w(\Lambda_i)]/\sqrt{\text{Det}[w(\Lambda_i)]} = \pm 1$, where Pf denotes the Pfaffian of unitary matrix $[w(\Lambda_i)]$ with components $w_{mn}(\mathbf{k}) = \langle u_m(\mathbf{k})|\Theta|u_n(-\mathbf{k})\rangle$. $u_m(\mathbf{k})$ are Bloch states for band $m$, $\Theta$ is the antiunitary time-reversal operator, and $\Lambda_i$ are the TRIM points in the BZ. The $Z_2$ invariant $\nu_0$ in 3D or $\nu$ in 2D is given by

$$(-1)^\nu = \prod_{i=1}^n \delta_i \qquad (1)$$

where $n$ is 4 in 2D and 8 in 3D. In 3D, the other three $Z_2$ invariants are obtained from partial products of sets of four $\delta_i$'s, similar to that of Eq. (1), corresponding to TRIM points lying in three independent planes of the BZ. In a crystal with inversion symmetry, Bloch wave functions are also eigenfunctions of the parity operator with eigenvalues $\xi_m(\Lambda_i) = \pm 1$ and the formula for $\delta_i$'s simplifies to (Fu, Kane, and Mele 2007):

$$\delta_i = \prod_m \xi_m(\Lambda_i), \qquad (2)$$

where the product is over the parities of pairs of occupied Kramer's doublets resulting from TRS at the TRIM points $\Lambda_i$ without multiplying the corresponding TR partners.

Fig. 2 shows the band structure of a 2D bilayer of Bi (Huang, Chuang, et al. 2013) along with parities of the three valence bands at the four TRIM points (one $\Gamma$ and three M) as an example of application of Eq. (2). From Fig. 2(a), $\delta_\Gamma = -1$, as the product of +1, -1, and +1. Similarly, $\delta_M = +1$, so that from Eq. (1), $\nu = 1$, and we have a nontrivial TI or a QSH state. The same computation for the bands of Fig. 2(b) yields $\nu = 0$ or the trivial state. The QSH state is seen to result from the band inversion at M, which is accompanied by a

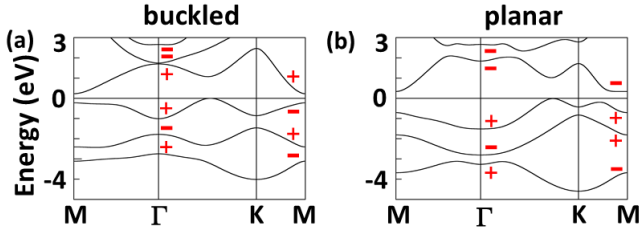

Fig. 2. Parities of bands at TRIM points in buckled (a) and planar (b) Bi thin films. Even (odd) parity is labeled + (-). Parity inversion between (a) and (b) is seen around the band gap at the M-point. (Huang, Chuang, et al. 2013)

change in the parity of the two states involved and thus changes the value of ν.

### IIB.1 Systems Without Inversion Symmetry

The preceding parity analysis is not applicable to systems without a center of inversion symmetry. However, in 2D, $Z_2$ can be cast in the general form (Fu and Kane 2006)

$$Z_2 = \frac{1}{2\pi} \left[ \oint_{d\tau} \boldsymbol{A}(\mathbf{k}) dl - \int_\tau F(\mathbf{k}) \, d\tau \right] \bmod 2, \quad (3)$$

where $\boldsymbol{A}(\mathbf{k}) = i \sum_{n=1}^{N} \langle u_n(\mathbf{k}) | \nabla_k u_n(\mathbf{k}) \rangle$ is the Berry connection (Berry 1984; Xiao, Chang, and Niu 2010) with sum over occupied states, and $F(\mathbf{k}) = \nabla_k \times \boldsymbol{A}(\mathbf{k})$ is the corresponding Berry curvature. The integrals are over half of the 2D BZ surface $\tau$ and its boundary $d\tau$. The 3D extension involves computing $Z_2$ from Eq. (3) for six different 2D tori obtained by taking various pairs of TRIM points in the 3D BZ. Several half-Heusler compounds have been investigated by Xiao et al. (Xiao, Yao, et al. 2010) along these lines using discrete **k**-meshes. (Fukui and Hatsugai 2007)

Berry connection $\boldsymbol{A}(\mathbf{k})$ and Berry curvature $F(\mathbf{k})$ in Eq. (3) are important quantities that can be constructed from Bloch wave functions, and enter the computation of topological properties of materials more generally. For example, anomalous contribution to Hall conductivity is given by appropriate line and surface integrals of $\boldsymbol{A}(\mathbf{k})$.(Haldane 2004) Also, magnetoelectric response can be obtained from a BZ integral over a Chern-Simons 3-form involving $A_\mu^{mn}(\mathbf{k}) = -i\langle u_m(\mathbf{k}) | \nabla_\mu | u_n(\mathbf{k}) \rangle$ and its derivatives. (Qi, Hughes, and Zhang 2008) First-principles computations of magnetoelectric response along these lines have been carried out in trivial and non-trivial insulators. (Coh et al. 2011)

### IIB.2 Interacting Systems

The single-particle Bloch states are not eigenstates in an interacting system, and therefore, the spectral function is no longer a δ-function but develops finite width as a function of $E$ for a given **k**, or equivalently, as a function of **k** for a given $E$. This is also the case in a disordered system due to disorder induced smearing of states.(Bansil et al. 1999; Bansil 1979a; Bansil 1979b; Schwartz and Bansil 1974; Stocks, Temmerman, and Gyorffy 1978) Using field theoretic methods $Z_2$ can be evaluated in terms of just the zero frequency one-particle Green's function $G(0,\mathbf{k})$ of the interacting system (Wang and Zhang 2012; Wang and Zhang 2013). In inversion symmetric crystals, the formulae become more tractable as $Z_2$ can be obtained from the parities of eigenvectors of the Hermitian matrices $G(0,\Lambda_i)$ at the TRIM points $\Lambda_i$ in the spirit of Eq. (2).

### IIC. Adiabatic Continuation Approach

Topological invariants are a manifestation of the overall geometry or curvature of the bulk system, and therefore two different systems with a similar bulk property or 'genus' are topologically equivalent.(Hasan and Kane 2010; Essin and Gurarie 2011) Fig. 3(a) shows surfaces of spherical and crumpled balls, both with the genus, $g = 0$, where the genus counts the number of holes.(Nakahara 1990) The doughnut and the coffee cup similarly have one hole or $g = 1$. The point is that we can obtain the genus of the crumpled ball from that of the spherical ball so long as we can smoothly or adiabatically deform one into the other without introducing holes. Similarly, changes in the Hamiltonian, which do not induce a band inversion anywhere in the BZ, will not change the value of $Z_2$. Examples of 'deformations' are: strains in the crystal structure; changes in the nuclear charges of constituent atoms while maintaining charge neutrality; or changes in the strength of the SOC. Adiabatic continuity arguments provide a powerful tool for connecting different topological families, and are used extensively for predicting new materials, especially systems without inversion symmetry by starting from a known topological insulator. (Lin, Wray, et al. 2010; Al-Sawai et al. 2010; Lin, Das, Wang, et al. 2013; Chadov et al. 2013). Fig. 4 gives an illustrative example of how HgTe can be connected to the non-centrosymmetric half-Heusler compound YPtSb as follows. Aside from changes in lattice constants, one first inserts Kr atoms into empty positions in the zinc-blende structure to obtain the hypothetical half-Heusler compound KrHgTe, where the inert Kr atoms hardly affect the low energy Hamiltonian. The nuclear charges on Kr, Hg and Te are then adjusted slowly by setting, $Z_{Kr} = 36+2x+y$, $Z_{Hg} = 80-2x$, and $Z_{Te} = 52-y$, and varying $x$ and $y$. In the phase space of Hamiltonians defined by the parameters $x$ and $y$, no band inversions are found numerically when one connects KrHgTe at $x=0$, $y=0$ to YPtSb at $x=1$, $y=1$, proving that YPtSb and HgTe are topologically equivalent (Lin, Wray, et al. 2010).

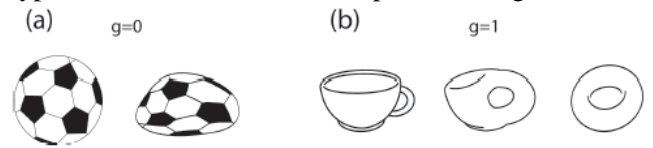

Fig. 3. Spherical and crumpled balls have the same genus of $g = 0$, in (a), but both the doughnut and the coffee cup with one hole in (b) are characterized by g = 1.

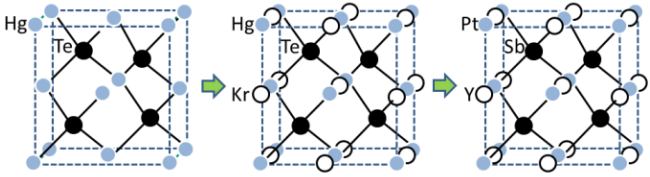

Fig.4: Adiabatic transformation of the known topologically nontrivial HgTe into the half-Heusler compound YPtSb (Lin, Wray, et al. 2010).

## IID. Surface/Edge State Computation

While the existence of gapless surface states on the interface of a TI with vacuum or another non-topological material is guaranteed by bulk-boundary correspondence considerations (Hasan and Kane 2010; Essin and Gurarie 2011), an actual computation must be carried out to ascertain the precise nature of these states. This can be done by considering a 2D slab of a 3D material or a 1D ribbon in 2D. (Fu, Kane, and Mele 2007; Fu and Kane 2007; Xia et al. 2009; Zhang, Liu, et al. 2009; Wang and Chiang 2014) Topological surface states must connect valence and conduction bands by crossing Fermi level ($E_F$) an odd number of times. The degeneracy of two surface bands at the TRIM points is protected by TRS, yielding linearly dispersing Dirac cones. The robustness of the gapless surface states has been demonstrated even when the dangling bond states dominate.(Lin, Das, Okada, et al. 2013) Topological surface states should be distinguished sharply from the well-known boundary states that arise in normal insulators with a long history in solid-state physics in that the latter type of states are less robust and can be removed by appropriate surface treatment. (Hasan and Kane 2010)

First-principles surface state calculations are computationally demanding, and the interpretation of results can be complicated by the spurious gaps resulting from interaction between the top and bottom surfaces of a finite slab. Fig. 5 gives an illustrative example of a first-principles surface state computation of a Γ-centered Dirac cone. The constant energy contours on the right side of the figure are seen to be circular with a helical spin-texture at 50 meV above the Dirac point, hexagonal at 150 meV, and snowflake like

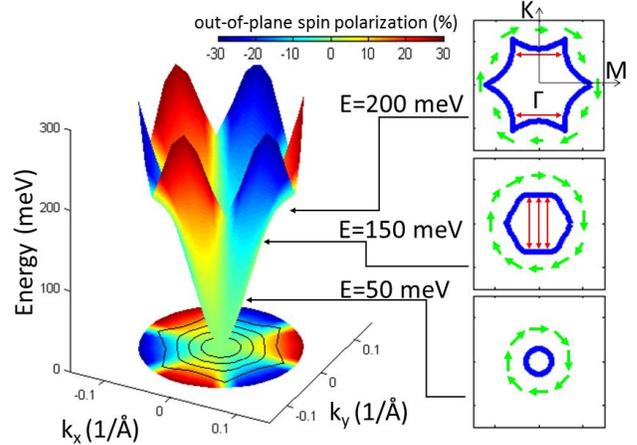

Fig. 5: First-principles surface state dispersion and out-of-the-plane spin polarization of the Dirac cone in $Bi_2Te_3$. Constant energy surfaces and the associated spin textures are shown at various energies above the Dirac point. (Hasan, Lin, and Bansil 2009)

at 200 meV with significant hexagonal warping and out-of-the-plane spin component. There also are in-plane deviations in that the spin direction is not always perpendicular to $\mathbf{k}$. (Fu 2009; Basak et al. 2011)

## IIE. Model Hamiltonians

A large variety of model Hamiltonians have been invoked for investigating 2D and 3D topological materials. For a few examples, see: Kane and Mele 2005a; Haldane 1988; Bernevig, Hughes, Zhang 2006; Fu 2009; Basak et al. 2011; Dzero et al. 2010; Zhang, Liu, Qi et al. 2009; Das and Balatsky 2013). Some are minimal and generic while others attempt more realism and material specificity by appealing to first-principles results, including efforts to construct localized Wannier type basis functions to reproduce bulk band structures. (Marzari and Vanderbilt 1997) Fig. 6 gives an example of how the electronic spectrum evolves with film thickness using the model of Das and Balatsky 2013), which builds the 3D crystal by stacking 2D bilayers (BLs) with opposite Rashba-type SOC in adjacent layers. Emergence of the topological Dirac cone at six BLs (6BL) is seen from the non-topological 1BL and 2BL films. Model Hamiltonians become unavoidable for addressing the superconducting state and other problems where first-principles treatment is not practical.

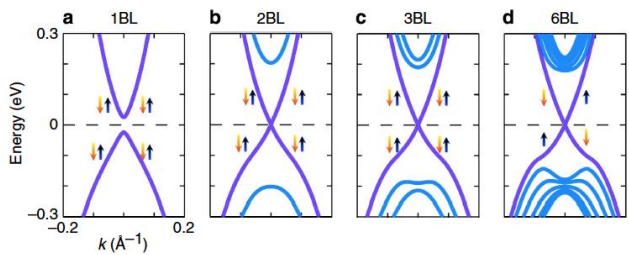

Fig. 6: Evolution of electronic spectrum as a function of number of bilayers (BL). (Das and Balatsky 2013)

## IIF. Topological properties and protections

We discuss now the extent to which the unique theoretically predicted helical spin textures and suppression of back scattering of topological surface states, dimensional crossover in thin films, and unusual magnetoelectric and Hall effects have held up to experimental verification.

Dirac cone dispersions and helical spin textures have been observed in ARPES experiments from a variety of topological materials in substantial accord with theoretical predictions, although the predicted position of $E_F$ with respect to the Dirac point is often not in agreement with measurements due to uncertain doping of naturally cleaved surfaces.(Xia et al. 2009) It has been demonstrated experimentally that the position of $E_F$ with respect to the Dirac point in the TIs can be manipulated via various dopings, which is important for reaching the topological transport regime, controlling carrier density and achieving n- or p-type doping for developing applications.(Hor et al. 2009; Hsieh, Xia, Qian, Wray, Dil, et al. 2009; Wray et al. 2010; Kong et al. 2011) Magnetic doping with Fe, Mn and Cr dopants has also been carried out. (Wray et al. 2011; Okada et al. 2011)

The predicted suppression of backscattering channels (Fu 2009; Zhou et al. 2009; Lee et al. 2009) has been observed in quasi-particle interference (QPI) patterns obtained from measured scanning tunneling spectra (Roushan et al. 2009). Moreover, as expected, the backscattering channels are seen to open up when magnetic impurities are introduced. (Okada et al. 2013) Although backscattering is the only channel available to 1D edge states, this is not the case for 2D surface states. Scattering in a 3D TI can be enhanced due to band structure effects on the Dirac cone states such as hexagonal warping and deviation of dispersion from linearity, and the mixing of various orbitals, which can also reduce the spin-polarization of surface states substantially. (Fu 2009; Basak et al. 2011) Resonances can be created by disorder and impurities as shown theoretically (Biswas and Balatsky 2009; Black-Schaffer and Balatsky 2013), and observed in scanning tunneling microscopy/spectroscopy (STM/STS) experiments. (Teague et al. 2012; Alpichshev et al. 2012a) Stronger localization can be induced by strong scalar disorder (Li et al. 2009; Guo et al. 2010) and magnetic impurities. (Biswas and Balatsky 2009; Alpichshev et al. 2012a)

Topological surface states can be gapped by finite size effects in thin films, due to interactions between the top and bottom surfaces. Such gapped Dirac cones have been shown to appear via ARPES and STM experiments in $Bi_2Se_3$ films of thickness less than 6 quintuple layers, (Zhang et al. 2010; Wang, Liu, et al. 2012; Neupane et al. 2014) where the critical thickness depends on the size of the bulk band gap. Notably, the 2D QSH phase is predicted to emerge in thin films of many 3D TIs. (Lin, Markiewicz, et al. 2010; Singh et al. 2013b; Liu et al. 2010; Ebihara et al. 2011)

Topological surface states exhibit novel magnetoelectric effects and non-abelian axion dynamics described by the term, $(\theta/4\pi)\mathbf{E}.\mathbf{B}$, in the electromagnetic action (Wilczek 1987). The axion field or axion angle $\theta$ is related to the magnetoelectric polarization $P_3 = \theta/2\pi$. (Qi, Hughes, and Zhang 2008) In a gapped surface state, changes in $\theta$ are associated with a surface Hall conductivity (Hasan and Kane 2010), which has been verified by Shubnikov de-Hass oscillation measurements in $Bi_2Te_3$ (Qu et al. 2010; Xu et al. 2014). Another interesting effect first predicted theoretically in $Bi_2Se_3$, $Bi_2Se_3$ and $Sb_2Te_3$ films doped with Cr and Fe is the quantum anomalous Hall effect (Yu et al. 2010), which has been observed in thin films of Cr-doped $(Bi,Sb)_2Te_3$(Chang et al. 2013). Heterostructure of a TI with a magnetic insulator can lead to a large magnetic gap in the topological surface state.(Luo and Qi 2013; Eremeev et al. 2013)

## III. Other topological states of quantum matter

Although this section discusses many prominent phases of quantum matter, a large number of other exotic possibilities exist in principle. (Thouless et al. 1982) The classifications can be based on various combinations of TRS, crystal space group symmetries, and particle-hole symmetry in superconductors. (Hasan and Kane 2010; Kitaev 2009; Schnyder et al. 2008; Fang, Gilbert, and Bernevig 2012; Slager et al. 2013; Chiu, Yao, and Ryu 2013; Hughes, Prodan, and Bernevig 2011; Shiozaki and Sato 2014)

## IIIA. Topological Crystalline Insulator (TCI)

In a TCI, spatial crystalline symmetries are the source of protection of topological states. (Fu 2011) TCIs are a natural extension of the $Z_2$ TIs in which topological states are protected by the TRS. The hallmark of a TCI is the existence of metallic surface states with novel characteristics on high-symmetry crystal faces. These surface states form a highly tunable 2D electron gas in which the band gap can be opened and tuned by external electric field or strain with potential applications in field-effect transistors, photodetectors, and nanoelectromechanical systems.(Liu, Hsieh, et al. 2014) SnTe and Pb$_{1-x}$Sn$_x$(Se,Te) alloys were the first TCI system driven by SOC predicted theoretically, and subsequently realized experimentally. (Hsieh et al. 2012; Tanaka et al. 2012; Dziawa et al. 2012; Xu, Liu, et al. 2012)

It is possible to realize a TCI without SOC. An example is the model of Fu (Fu 2011) in which topological surface states are protected by the combination of TRS and point-group symmetries for spinless fermions. For analyzing the topological invariant, note that for spinless fermions, $(TU)^2 = -1$ in a TCI, where $U$ is the unitary operator for the point group symmetry operation, and $T$ is the anti-unitary time-reversal operator. This is equivalent to the time-reversal operator $\Theta$ for spinfull fermions. Therefore, although $T$ itself does not guarantee two-fold degeneracy, the combination $TU$ for $C_4$-symmetry gives four-fold degeneracy at the four momentum points: $\Gamma = (0, 0, 0)$; M $= (\pi, \pi, 0)$; Z $= (0, 0, \pi)$; and, A $= (\pi, \pi, \pi)$. The topological invariant $\nu_0$ is given by

$$(-1)^{\nu_0} = \delta_{\Gamma M} \delta_{AZ}, \tag{4}$$

where

$$\delta_{k_1 k_2} = e^{i \int_{k_1}^{k_2} dk.A_k} \frac{\mathrm{Pf}[w(k_2)]}{\mathrm{Pf}[w(k_1)]} \tag{5}$$

in terms of the Berry connection $A_k$, see Eq. (3), and Pfaffians of anti-symmetric matrices $w(k_i)$, where $w_{mn}(k_i) = \langle u_m(k_i)|UT|u_n(-k_i)\rangle$. The line integrals are between the points $k_1$ and $k_2$ in the corresponding 2D planes.

### IIIB. Disorder or interaction driven TIs

Anderson's pioneering work showed that disorder can lead to a metal insulator transition. (Anderson 1958) For a variety of tight-binding models, the existence of a disorder induced inverted insulating gap has been shown in 2D as well as 3D systems, and supported by the observation of quantized conductance when the $E_F$ lies in the gap. (Li et al. 2009; Guo et al. 2010)

An axion insulator has a quantized magnetoelectric response given by the axion angle θ, which assumes a quantized value of π in the topological state. (Essin, Moore, and Vanderbilt 2009) For a commensurate antiferromagnetic insulator, the θ=π state can be obtained through the combination of time-reversal and lattice translational symmetries (Mong, Essin, and Moore 2010). Axion insulator state is theoretically predicted in magnetic systems such as the iridates (Wan et al. 2011) and osmium compounds in the geometrically frustrated spinel structure (Wan, Vishwanath, and Savrasov 2012). Since TRS is broken, axion insulators lack protected surface states.

Quantum anomalous Hall (QAH) insulator is another example of a TRS breaking state in which band inversion between the majority and minority spin states is driven by the magnetic exchange energy. This state is theoretically predicted in a number of models, following the original proposal of Haldane with bond currents on a honeycomb lattice (Haldane 1988). These models include, localization of band electrons(Onoda and Nagaosa 2003), a two-band model of a 2D magnetic insulator (Qi, Wu, and Zhang 2006), and magnetically doped topological (crystalline) insulator thin films (Yu et al. 2010; Fang, Gilbert, and Bernevig 2014; Wang et al. 2013b) or even trivial thin films. (Doung et al. 2015)

### IIIC. Topological superconductors

Form of the low-energy Hamiltonian of a fully gapped superconductor is similar to that of a TI in many ways. A topological superconductor is obtained when the bulk system has a pairing gap, but supports gapless Majorana modes at the boundary. (Roy and Kallin 2008) For instance, the TRS breaking (chiral $p$+i$p$) and TRS preserving $p$±i$p$ pairing states are analogous to the integer quantum Hall and quantum spin-Hall states, respectively. The former case supports chiral propagating Majorana edge modes, while the latter supports the counter-propagating Majorana edge modes (Elliott and Franz 2014) which are topologically protected against time-reversal invariant perturbations.

Chiral superconductivity has been predicted in Sr$_2$RuO$_4$ (Mackenzie and Maeno 2003), doped graphene (Pathak, Shenoy, and Baskaran 2010); Nandkishore, Levitov, and Chubukov 2012), and other systems.(Liu, Liu, et al. 2013; Hsu and Chakravarty 2014) The B-phase of $^3$He may harbor a time-reversal-invariant topological superfluid state, although the corresponding surface states are yet to be identified. (Chung and Zhang 2009; Qi and Zhang 2011) Superconductivity seen experimentally by doping Bi$_2$Se$_3$ with Cu (Wray et al. 2010) is argued to be the signature of parity-odd superconducting state. (Sato 2009; Fu and Berg 2010) A fully gapped state is also observed in In-doped TCI SnTe (Novak et al. 2013). Fu and Kane (Fu and Kane 2008) propose an alternative approach in which the topological state is induced via proximity effects between the topological and trivial $s$- wave superconductors. This proximity effect has been observed in $Bi_2Se_3$ on a $NbSe_2$ conventional superconductor substrate (Wang, Liu, et al. 2012) as well as an unconventional $d$-wave $Bi_2Sr_2CaCu_2O_{8+\delta}$ superconducting substrate (Wang, Ding, et al. 2013). A new type of topological mirror superconductor has been predicted (Zhang, Kane, and Mele 2013; Tsutsumi et al. 2013) where the protection is through the combination of mirror and time-reversal symmetries like a TCI.

A number of proposals have been made for realizing Majorana state,(Elliott and Franz 2014) leading to the general principle that Majorana edge modes are inherited by the ends of a one-dimensional chain of magnetic impurities or adatoms or a helical Shiba chain on a superconducting substrate. (Pientka, Glazman, and von Oppen 2013; Heimes, Kotetes, and Schön 2014) Such bond states have been reported by STM in Fe chains on superconducting Pb.(Nadj-Perge et al. 2014)

### IIID. Weyl and 3D Dirac Semimetal Phase

Gapless cone-like dispersions can exist in the 3D bulk electronic spectrum described by the Hamiltonian, $H(\boldsymbol{k}) = v_{ij}k_i\sigma_j$, which is similar to the Weyl equation (Weyl 1929), with associated Chern number given by sgn(det[$v_{ij}$])=±1. If the TRS or the inversion symmetry is broken, these cones can become nondegenerate except at the (Weyl) nodes or points, yielding a topologically protected Weyl semimetal phase. Weyl nodes can be created or annihilated only when two nodes with opposite signs of Chern numbers come together (Vafek and Vishwanath 2014). The total number of Weyl nodes can be shown to come in multiples of 4 for TRS preserved and 2 for TRS broken systems. (Ojanen 2013; Burkov and Balents 2011a) Weyl semimetal phase may sometimes be viewed as a bridge between a $Z_2$ TI with broken inversion symmetry and a trivial band insulator (Murakami 2007), or alternatively, as a bridge between an axion insulator with broken TRS and a trivial Mott insulator (Wan et al. 2011). It is possible to have two degenerate Weyl points with the same sign of Chern numbers or to have multiple degenerate Weyl points protected by point group symmetry. (Fang, Gilbert, and Bernevig 2012)

Spin degenerate 3D Dirac cones, which consist of two gapless Weyl nodes, are also possible. Accidental gapless Dirac cones can occur at the topological phase transition between a TI and a trivial insulator if both inversion and time-reversal symmetries are intact, (Xu, Xia, et al. 2011) although cases where the Dirac nodes are protected by space-group symmetry are more

interesting.(Young et al. 2012; Wang, Sun, et al. 2012; Wang et al. 2013b)

Weyl and 3D Dirac semimetal phases would have important applications including the realization of anomalous Hall effect (Yang, Lu, and Ran 2011; Burkov and Balents 2011b; Xu, Xia, et al. 2011), nontrivial electromagnetic responses (Turner and Vishwanath 2013), Majorana excitations (Meng and Balents 2012), a disconnected, yet protected, Fermi surface (FS) or a 'Fermi arc' (Wan et al. 2011), and possible Weyl superconductors.(Meng and Balents 2012)

### IIIE. Topological Phase Transition (TPT)

A TPT, which is driven by changes in the topology of bulk band structure, is very different from the familiar phase transitions such as the melting of a solid, which are characterized by broken symmetries and sharp anomalies in thermodynamic properties. Many theoretical studies show that TPTs can be induced by tuning the band structure using chemical substitution, strain, or pressure, or via electron correlation effects. (Xu, Xia, et al. 2011; Sato 2009; Wray et al. 2011; Wu, Brahlek et al. 2013; Wan et al. 2011; Pesin and Balents 2010)

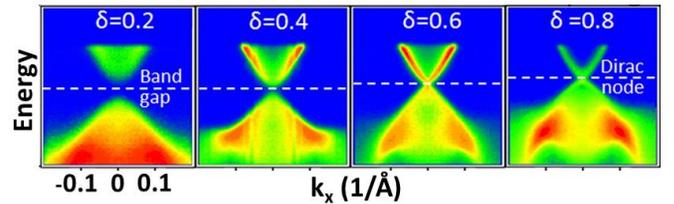

Fig. 7: ARPES spectra of $TlBi(S_{1-\delta}Se_\delta)_2$ showing a TPT as $\delta$ increases from 0.2 to 0.8. (Xu et al. 2011).

TPTs have been observed experimentally in $TlBi(S_{1-\delta}Se_\delta)_2$ (Xu et al. 2011; Sato et al. 2011) as well as in the TCI $Pb_{1-x}Sn_xSe$. (Dziawa et al. 2012; Zeljkovic, Okada, et al. 2015) An example is given in Fig. 7, which shows the evolution of the spectrum in $TlBi(S_{1-\delta}Se_\delta)_2$ with Se content $\delta$. The band gap is trivial at $\delta$=0.2, closes around $\delta$=0.6, and reopens at $\delta$=0.8 with the appearance of linearly dispersing bands connecting valence and conduction bands. Similar results are found in a time-domain terahertz spectroscopy study of $(Bi_{1-x}In_x)_2Se_3$ (Wu, Brahlek et al. 2013). TPTs can be induced via laser or microwave pumping to produce a non-equilibrium topological state or a Floquet TI. (Lindner, Refael, and Galitski 2011; Kitagawa et al. 2010; Gu et al. 2011; Dóra et al. 2012; Katan and Podolsky 2013; Rechtsman et al. 2013; Kundu and Seradjeh 2013; Wang, Steinberg, et al. 2013; Wang, Wang, et al. 2014; Perez-Piskunow et al. 2014)

## IV. Survey of Topological Materials

### IVA. Materials discovery strategies

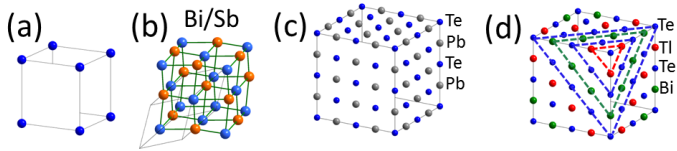

Fig. 8: Simple cubic (a) and the related pseudo-cubic rhombohedral structure. Rocksalt (c) and the related pseudo-rocksalt rhombohedral structure (d).

A key ingredient for realizing a topologically non-trivial state is the presence of an SOC driven band inversion as discussed in Sec. IIA. Therefore, natural candidates for the topological materials discovery process are traditional semiconductors and insulators containing heavy elements. The size of the band gap here is not important because it could be inverted and/or extended throughout the BZ via controls of strain, alloying and/or confinement, so that semimetals also provide viable base materials. For example, a Bi/Sb alloy was the first materials realization of a 3D TI in which the parent Sb is a semimetal with a nontrivial band topology. (Fu and Kane 2007; Hsieh et al. 2008; Hsieh, Xia, Wray, Qian, Pal, et al. 2009) In non-magnetic compounds, the search for new topological materials should typically start thus by considering systems with an even number of electrons per unit cell, which could in principle be accommodated in completely filled bands. [This of course does not apply to magnetic or strongly correlated systems.] Large gap ionic compounds in which the conduction and valence bands are formed from distinct atomic orbitals can be ruled out. Intermetallics and covalent bonded materials are more appropriate. Since strong SOC resides in the bottom part of the periodic table, by excluding metallic Pb and Group 3A elements, we arrive at Sn in Group 4A and Bi/Sb in Group 5A, which are not too ionic, as the elemental ingredients of choice for creating new topological materials.

The crystal structure is also a key player in controlling band topology. For example, a hypothetical simple cubic Sb or Bi crystal (Fig. 8a) will be metallic as it contains an odd number of valence electrons (3 or 5 in group 5A). Sb and Bi however occur naturally in a rhombohedral lattice with two basis atoms, which can be viewed as a pseudocubic structure (Hofmann 2006), Fig. 8b, yielding an even number of electrons in the unit cell, and opening up the possibility of an insulating phase. An even number of electrons in the unit cell is maintained when we replace Sb by Sn and Te in alternate layers since Sn and Te lie on the left and right hand sides of Sb in the periodic table, giving a rocksalt type face-centered-cubic (fcc) lattice with two basis atoms (Fig. 8c). This strategy has been useful in manipulating $Pb_{1-x}Sn_xTe$, which undergoes inversions at four equivalent L-points in the BZ with increasing $x$, and thus remains $Z_2$ trivial for all $x$. [Sec. IVB.6 below discusses how a TCI phase emerges in $Pb_{1-x}Sn_xTe$.] But, a strong TI can be realized via a rhombohedral distortion along [111], which limits band inversion to just one of the four L-points. This approach successfully predicted $TlBiTe_2$ with pseudo PbTe structure to be a TI where the rhombohedral distortion is induced chemically by replacing alternate layers of Pb by Tl and Bi (Fig. 8d). (Lin, Markiewicz, et al. 2010; Hsieh et al. 2012)

Note that out of a total of 230 available crystallographic space groups, the three TI families of $TlBiSe_2$, and distorted (Pb/Sn)Te and (Bi/Sb) alloys belong to the same space group #166 (R-3m), which transforms into #164 (P-3m1) when a conventional hexagonal cell is considered. $Bi_2Se_3$, ternary tetradymites, $Ge_mBi_{2n}Te_{m+3n}$ series, $(Bi_2)_m(Bi_2Te_3)_n$, and BiTeCl TIs all belong to the same or similar space groups. Interestingly, their structures can be viewed as being more or less pseudo-cubic with most atoms having essentially a coordination of 6. We may thus consider Bi/Sb to be a prototype TI for a variety of Bi/Sb based layered TIs as well as for the (Pb/Sn)Te alloys. In contrast, grey Sn, which is connected adiabatically to several zinc-blende type TIs, presents a second distinct structural branch favored by topological materials. These observations point to the value of space groups in searching for new topological materials. Yet other classes of TIs involving $d$ and $f$ electrons highlight the role of strong electron correlations in producing new topological phases. Notably, some topological phases are not driven by the SOC.(Fu 2011; Alexandradinata et al. 2014)

Standard band theory techniques can be deployed for treating the electronic structure and topological properties of 2D films, see Sec. IVD below, by constructing an effective 3D 'crystal' obtained by stacking replicas of the 2D film separated by vacuum layers. While weakly correlated materials can be modeled in a parameter free manner within the band theory framework, this is generally not the case when electronic correlations are strong, and a variety of parameters are often invoked in order to make a headway (e.g. value of U in an LDA+U computation or the strength of the SOC); we refer to the discussions of Sections IVE-IVI and the cited literature below for the specifics in various cases.

## IVB. Bi/Sb-based Materials

### IVB.1 $Bi_{1-x}Sb_x$: First 3D TI

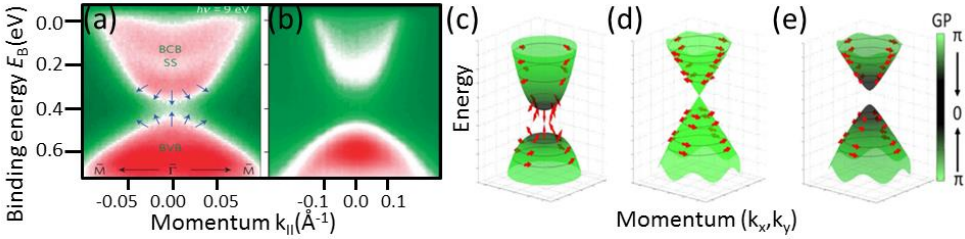

Fig. 9: Band dispersion via ARPES in magnetically doped Bi₂Se₃ (a) and its 3QL thick film (b). Schematic spin-texture for a TRS breaking gapped Dirac cone (c), a gapless Dirac cone (d), and a gapped Dirac cone (e). (Xu, Neupane, et al. 2012)

Although Bi and Sb are both semimetals, the ordering of conduction and valence bands in Bi and Sb at the three L-points in the rhombohedral BZ is different. In Bi₁₋ₓSbₓ alloys, as $x$ increases, band gaps close and reopen at the three L-points with a critical point at $x \approx$ 4%, and the system becomes a direct-gap semiconductor at $x \approx$ 8%. These considerations led to the theoretical prediction that Bi-Sb alloys harbor a TI phase (Fu and Kane 2007), and to the experimental discovery of the first 3D TI in Bi₀.₉Sb₀.₁, which was established by observing the hallmark topological surface states via ARPES on the [111] surface (Hsieh et al. 2008). Gapless surface states have been observed on the [110] surface as well (Zhu, Stensgaard, et al. 2013). Also, the QPI patterns in STS experiments show the expected suppression of the backscattering channels, while scattering channels between different pieces of the FSs remain open. (Roushan et al. 2009; Gomes et al. 2009) Interestingly, STS experiments show Landau levels as well as QPI patterns of nontrivial surface states in the topological semimetal Sb (Soumyanarayanan et al. 2013).

## IVB. 2 Bi₂Se₃, Bi₂Te₃, Sb₂Te₃: 2nd generation TIs

Bi₂Se₃, Bi₂Te₃, and Sb₂Te₃ are strong TIs, which share a rhombohedral structure containing blocks of quintuple layers (QLs). Non-trivial topology results from band inversions driven by SOC in the $p$-orbital manifold at the Γ-point. The band gap in Bi₂Se₃ is as large as 0.3 eV. Unlike the multiple surface states in Bi₁₋ₓSbₓ, these materials possess the advantage of exhibiting only a single Dirac cone on their naturally cleaved [111] surfaces. (Xia et al. 2009; Hsieh, Xia, Qian, Wray, Dil, et al. 2009; Zhang, Liu, et al. 2009; Hsieh, Xia, Qian, Wray, Meier, et al. 2009; Chen et al. 2009) As a result, they have become the workhorse materials of the field with many ARPES, (Hsieh, Xia, Qian, Wray, Dil, et al. 2009; Xia et al. 2009; Chen et al. 2009) STM, (Hor et al. 2009; Okada et al. 2011; Alpichshev et al. 2012b; Zhang, Cheng, et al. 2009) transport, (Qu et al. 2010; Xiong et al. 2012) and optical (Valdes Aguilar et al. 2012; Wu, Brahlek et al. 2013; Jenkins et al. 2010; Jenkins et al. 2012) studies.

An experimental challenge has been to practically realize the bulk insulating state and to manipulate the position of the $E_F$. Since bulk Bi₂Se₃ is an n-type semiconductor, many studies attempt doping with extra holes. 0.25% Ca doping and NO₂ surface deposition tunes $E_F$ to the Dirac point and fully removes the bulk conducting band from $E_F$, (Hsieh, Xia, Qian, Wray, Dil, et al. 2009; Hor et al. 2009)while Sb doping in Bi₂Te₃ (Kong et al. 2011; Zhang, Chang, et al. 2011) and Bi₂Se₃ (Analytis et al. 2010) has been shown to control the carrier density and $E_F$. In (Bi₁₋ₓSbₓ)₂Te₃ alloy, increasing Sb content shifts $E_F$ down from n- to p-type regime. On a nanotemplate of this sample, ambipolar gating effects have been reported. (Kong et al. 2011; Chen, Qin, et al. 2010; Steinberg et al. 2010) In-doped (Bi₁₋ₓInₓ)₂Se₃ films exhibit a metal insulator transition. (Wu, Brahlek, et al. 2013; Brahlek et al. 2012) DFT calculations predict that the position of the topological surface state can be tuned in heterostructures of a TI with various band insulators. (Wu, Chen, et al. 2013a; Menshchikova et al. 2013; Zhang, Zhang, Zhu, et al. 2012)

Doping with magnetic impurities is interesting as it gives insight into the effects of TRS breaking perturbations.(Okada et al. 2011; Schlenk et al. 2012; Jiang, Li, et al. 2013) The opening of a magnetic gap at the Dirac node is suggested by ARPES measurements on Fe/Mn doped Bi₂Se₃.(Chen, Chu, et al. 2010) However, other factors such as spatial fluctuations and surface chemical disorder (Zhang, Chang, et al. 2011; Beidenkopf et al. 2011) could be responsible since a gap-like feature is observed also in doped non-magnetic samples. (Xu, Neupane, et al. 2012) Spin-resolved ARPES reveals a hedgehog spin texture of the gapped Dirac cones in Mn-doped Bi₂Se₃, Figs. 9(a,c), which is distinct from that of a gapped Dirac cone due to confinement effects in thin films, Figs. 9(b,e). Finally, the long-sought quantum anomalous Hall effect has been demonstrated in (Bi,Sb)₂Te₃ thin films, (Chang et al. 2013; Zhang, Chang, et al. 2013) where the surface state tunneling gap in the undoped system can be closed/reopened via Cr doping, and a quantized Hall signal of the expected value is seen.

Topological superconductivity could be induced via doping or pressure, although this has not been confirmed experimentally. In Cu$_x$Bi$_2$Se$_3$, the transition temperature $T_c$ goes up to 3.8 K with Cu doping, (Wray et al. 2010) and superconductivity occurs through an unusual doping mechanism in that the spin-polarized topological surface states remain intact at the $E_F$. Low-temperature electrical resistivity and Hall effect measurements (Kirshenbaum et al. 2013; Zhang, Zhang, Yu, et al. 2012) on Bi$_2$Se$_3$ and Bi$_2$Te$_3$ single crystals under pressures ≤50 GPa show the onset of superconductivity above 11 GPa. $T_c$ and the upper critical field $H_{c2}$ both increase with pressure up to 30 GPa, where they peak with values of 7 K and 4 T, respectively. With further increase in pressure, $T_c$ remains anomalously flat even though the carrier concentration increases tenfold, pointing to an unusual pressure induced topological pairing state in Bi$_2$Se$_3$.

### IVB. 3 Ternary tetradymites, Ge$_m$Bi$_{2n}$Te$_{(m+3n)}$ series

The large family of tetradymite-like layered TIs with formulae $B_2X_2X'$, $AB_2X_4$, $A_2B_2X_5$, and $AB_4X_7$ ($A$ = Pb, Sn, Ge; $B$ = Bi, Sb; $X$, $X'$ = S, Se, Te) offers substantially greater chemical and materials tunability compared to its binary cousins discussed above. Many of these compounds have been synthesized and ARPES (Xu et al. 2010; Neupane et al. 2012; Zhang, Chang, Zhang, et al. 2011; Eremeev, Landolt, et al. 2012; Okamoto et al. 2012) and transport measurements (Xiong et al. 2012; Taskin et al. 2011; Ren et al. 2010) are available. Theoretically predicted single surface Dirac cones have been observed in several of these materials via ARPES as well as pump-probe spectroscopy of unoccupied surface states.(Niesner et al. 2012; Niesner et al. 2014)

The crystals of this series are built by stacking layers. We highlight the materials flexibility by considering the example of Ge$_m$Bi$_{2n}$Te$_{(m+3n)}$. It reduces to Bi$_2$Te$_3$ for $m$=0 and $n$=1. By increasing $m$, we get GeBi$_2$Te$_4$, which can be viewed as an insertion of a GeTe layer into Bi$_2$Te$_3$, yielding 7-layer blocks. An increase in $n$ now gives GeBi$_4$Te$_7$ with alternating stacks of 5-layer blocks of Bi$_2$Te$_3$ and 7-layer blocks of GeBi$_2$Te$_4$, which resembles a heterostructure. Many variations in stacking and composition can thus be made to tune properties of the surface Dirac cone. (Eremeev, Landolt, et al. 2012) The bulk gap varies over a wide range from 0 to 0.5 eV. A fully isolated Dirac cone at the $E_F$ can be obtained in Sb$_x$Bi$_{2-x}$Se$_2$Te for $x$=1.67. (Neupane et al. 2012)

### IVB. 4 Thallium Based Ternary Chalcogenides

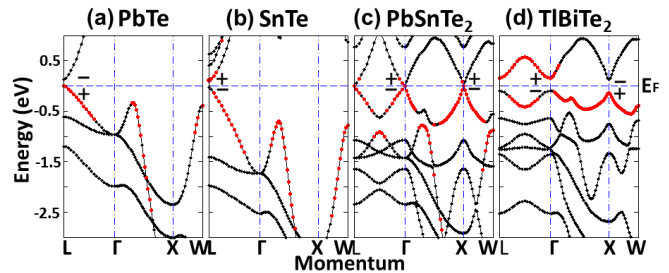

Fig. 10: Band structures of fcc PbTe (a) and SnTe (b), and rhombohedral PbSnTe$_2$ (c) and TlBiTe$_2$ (d), all plotted in the BZ of an fcc lattice. Parities at TRIM points are shown. Size of red dots is proportional to the weight of the s-orbital on the M atoms. (Lin, Markiewicz, et al. 2010)

Although Pb and Sn based chalcogenides were studied extensively in the 1980s in search of Dirac fermions, (Fradkin, Dagotto and Boyanovsky 1986) Tl-based III-V-VI$_2$ chalcogenides MM'X$_2$ [M = Tl, M' = Bi or Sb, and X = Te, Se, or S] have been recognized as TIs very recently. Their rhombohedral lattices, see Fig. 8(d), can be embedded in a 2×2 supercell of the fcc lattice. Fig. 10 gives insight into how non-trivial phases evolve from the trivial band structures of PbTe and SnTe.(Lin, Markiewicz, et al. 2010; Hsieh et al. 2012; Safaei, Kacman, and Buczko 2013) The analysis is simplified greatly by choosing the M-atom (Pb, Sn, or Tl), which is a center of inversion symmetry, to lie at the origin of the real space lattice. Bands are plotted using black and red dots where size of the red dots is proportional to the weight of the s-orbital on the M atom. This representation is easily generated in band structure codes, and allows a straightforward analysis of parities. Since the s-orbital on the M atom is even, its weight must be strictly zero for a state with odd parity. The parities of wave functions at TRIM points in the BZ are thus easily obtained from colors of the dots, being red for even and black for odd parity, and changes in the Z$_2$ invariants of the band structure can be monitored simply by counting the number of inversions of red and black dots at TRIM points around the $E_F$.

In order to understand how the topology of bands evolves in Fig. 10, it is useful to plot all bands in the BZ of the fcc lattice, so that bands in Figs. 10(c) and (d) for rhombohedral lattices refer to an extended zone scheme. (Lin, Markiewicz, et al. 2010) The top of the valence bands in PbTe lies at the L-points with a small gap, Fig. 10(a). The band structure of the hypothetical ternary compound PbSnTe$_2$ obtained by replacing one of the Pb atoms by Sn in the supercell is shown in Fig. 10(c). Pb-to-Sn replacement is seen to induce band inversions at both the Γ- and the three X-points around $E_F$. Because the total number of inversions is even, as is the case in SnTe at the four L-points in Fig. 10(b), there

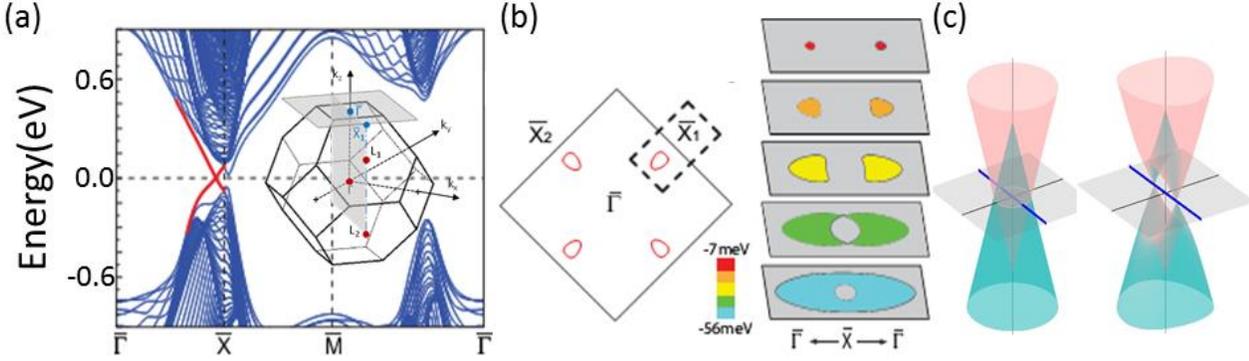

Fig. 11: [001] surface states of SnTe. (a) Surface (red lines) and bulk (blue lines) bands. BZ is shown in the inset. (b) FS (left) and a set of constant energy surfaces showing the Lifshitz transition (right). (c) Model of two non-interacting (left) and interacting (right) Dirac cones. (Hsieh et al. 2012; Wang, Tsai, et al. 2013)

is no change in $Z_2$, and PbSnTe$_2$ remains topologically trivial like PbTe. In fact, the band inversion in going from PbTe to PbSnTe$_2$ can be adduced to occur first at $\Gamma$ and then at X, suggesting that a small rhombohedral lattice distortion or chemical tuning by replacement of Pb and Sn with other elements, could prevent the band inversion associated with the three X-points, and produce a non-trivial phase. This indeed is seen to be the case in Fig. 10(d) when Pb is replaced by Tl and Sn by Bi.

Band theory calculations predict (Yan, Liu, et al. 2010; Lin, Markiewicz, et al. 2010; Eremeev, Koroteev, and Chulkov 2010; Eremeev et al. 2011) that four Tl based chalcogenides MM'X$_2$ [M=Tl, M' = Bi/Sb, and X = Te/Se] are topologically nontrivial. These theoretical predictions were verified experimentally soon after they were made by observing single-Dirac-cone surface states in TlBiTe$_2$ and TlBiSe$_2$.(Sato et al. 2010; Chen, Liu, et al. 2010; Kuroda, Ye, et al. 2010) Notably, TlBiTe$_2$ could be a topological superconductor.(Chen, Liu, et al. 2010; Yan, Liu, et al. 2010) On the other hand, TlSbS$_2$ does not crystallize in the rhombohedral structure, but the trivial TlBiS$_2$ does. The correct prediction of different topological phases between two isostructural compounds (TlBiS$_2$ and TlBiSe$_2$) suggests the possibility of a topological phase transition in TlBi(Se,S)$_2$ alloys, which has been realized experimentally, (Xu et al. 2011) see Section IIIE above. On the other hand, because the structure of Sb$_2$Se$_3$ is different from Bi$_2$Se$_3$ or Sb$_2$Te$_3$, it will be difficult to realize such a transition in (Bi,Sb)$_2$Se$_3$ or Sb$_2$(Se,Te)$_3$ alloys.

## IVB. 5 Non-centrosymmetric Bi-compounds

Bi-based non-centrosymmetric compounds BiTeX (X=Cl, I, Br) show the presence of giant Rashba-type SOC.(Ishizaka et al. 2011; Bahramy, Arita, and Nagaosa 2011; Eremeev, Nechaev, et al. 2012) Band structure computations on BiTeCl find oppositely polarized layers of Bi and Te (Chen et al. 2013) so that the system resembles a topological p-n junction. A single V-shaped Dirac cone is reported via ARPES on the n-type, but not the p-type surface in BiTeCl. (Chen et al. 2013) The nature of surface states however remains uncertain with DFT predicting BiTeCl to be a trivial insulator, (Eremeev, Nechaev, et al. 2012; Landolt et al. 2013) although BiTeI is predicted to become a non-trivial TI under pressure. (Bahramy, Yang, et al. 2012)

## IVB. 6 First topological crystalline insulator (TCI): (Pb,Sn)Te

As discussed in Sec. IIIA, surface states in a TCI are protected by the combined effects of time-reversal and crystal symmetries. A TCI supports an even number of Dirac cones and band inversions in sharp contrast to a TI. (Pb/Sn)Te and related compounds were first predicted theoretically to host the TCI phase. (Hsieh et al. 2012) The experimental realization followed almost immediately. (Tanaka et al. 2012; Dziawa et al. 2012; Xu, Liu, et al. 2012) To date this remains the only known TCI family verified by experiments.

Crystal structure of this system is based on the rocksalt fcc structure, Fig. 8(c). The band gaps at four equivalent L-points in the BZ in SnTe are inverted relative to PbTe, Figs. 10 (a,b). The band gap in Pb$_{1-x}$Sn$_x$Te alloys closes/reopens with an even number of inversions between the two end compounds, so that even though neither SnTe nor PbTe can be a $Z_2$ TI, we obtain a TCI phase driven by the mirror symmetry of the fcc lattice. Note that standard parity analysis cannot be used here for the identification of an intrinsic band inversion because parity eigenstates depend on the choice of origin. For example, if the inversion center is shifted from Pb or Sn to a Te atom, the states at L would change parity but those at $\Gamma$ will not. Obviously, one cannot identify a TCI by comparing the sequence of parity eigenstates at L and Γ. The trick of plotting the weight of the key orbital involved in inversion discussed in connection with the red dots in Fig. 10 can however still be used. Here the relevant orbital is the Te *p*-orbital, which gets inverted in going from PbTe to SnTe. Specifically, SnTe at ambient pressure is a TCI with mirror Chern number $n_M = -2$, where the non-zero value of $n_M$ indicates the existence of surface states on any crystal surface symmetric about the {110} mirror planes.

The practical situation in SnTe is complicated by the rhombohedral distortion of the structure, (Iizumi et al. 1975) which breaks the mirror symmetry to produce gapped surface Dirac cones. (Hsieh et al. 2012) Moreover, SnTe surface is heavily p-doped by naturally occurring Sn vacancies, which lower the chemical potential below the bulk valence band maximum, (Burke et al. 1965) and push the Dirac cone into the valence bands. (Littlewood et al. 2010) But, this p-doping is absent in Pb-rich samples, (Takafuji and Narita 1982) and indeed, in topological compositions of Pb$_{1-x}$Sn$_x$Te the Dirac cone is seen via ARPES, and its expected spin structure is verified by spin-resolved ARPES experiments. (Xu, Liu, et al. 2012)

The commonly studied surface is the {001} surface where the $\Gamma L_1 L_2$ plane of the bulk BZ projects onto the $\Gamma X_1$ symmetry line in the surface BZ with both $L_1$ and $L_2$ projecting onto $X_1$, see inset in Fig. 11(a). The mirror Chern number $n_M = -2$ dictates the existence of two pairs of counter-propagating, spin-polarized surface states with opposite mirror eigenvalues along the $X_1$ - $\Gamma$ - $X_1$ line, replicated along the $\overline{X}_2-\Gamma-\overline{X}_2$ line via rotational symmetry. We thus obtain four Dirac points located on the four equivalent $\Gamma X$ lines. As E$_F$ decreases from the Dirac point, the FS initially consists of two disconnected hole pockets away from $X$, which subsequently reconnect to form a large hole and a small electron pocket, both centered at $X$, undergoing a Lifshitz transition in FS topology as depicted in Fig. 11(b). Role of surface passivations in destroying trivial surface states on the (111) polar surface of SnTe has been discussed. (Eremeev et al. 2014)

The complicated surface band structure and spin-textures discussed in the preceding paragraph can be understood in a model involving two coaxial Dirac cones where one starts with two non-interacting "parent" Dirac cones centered at X, which are vertically offset in energy, Fig. 11(c). Hybridization between these two parent cones opens a gap at all points except along the mirror line, leading to a pair of lower-energy "child" Dirac points away from X. The parent Dirac nodes are protected by TRS and cannot be gapped by removing the mirror symmetry, but the child Dirac nodes can be gapped by breaking mirror symmetry. Notably, the two parent Dirac cones must have different orbital characters since they belong to different eigenvalues of the mirror operation. In SnTe, the lower (higher) parent Dirac cone is primarily composed of Sn-p$_z$ (Te-p$_x$) orbital around the (π,0) point. Wang, Tsai et al. (2013) present a 4×4 model Hamiltonian along these lines based on up- and down-spin Sn-p$_z$ and Te-p$_x$ orbitals, which captures salient features of the corresponding first-principles surface states and their spin-textures.

The aforementioned orbital textures would be expected to yield intensity asymmetries in the QPI patterns obtained from STS spectra. The scattering between the p$_z$-like hole branch will be strong while that between different orbitals on the electron branch will be suppressed. Also, if mirror symmetry is broken along only one of the two mirror planes, then we will obtain massive Dirac cones along one direction, while the Dirac cones along the other direction will remain massless. Such a coexistence of massive and massless Dirac cones has been adduced via the observation of three non-dispersive features in the STS spectra, including the mapping of the associated dispersions in substantial accord with theoretical predictions. (Okada et al. 2013; Zeljkovic et al. 2014) Experiment and theory should however be compared at the level of spectral intensities including matrix element effects, which are important in STS, ARPES, and other highly resolved spectroscopies. (Nieminen et al. 2009; Nieminen et al. 2012; Bansil and Lindroos 1999; Sahrakorpi et al. 2005; Lindroos and Bansil 1996; Bansil et al. 2005; Huotari et al. 2000; Mijnarends and Bansil 1976; Smedskjaer et al. 1991; Mader et al. 1976)

The surface of a TCI provides an especially rich sandbox for exploring how spin and orbital textures play out in the presence of many different types of carriers and van Hove singularities in the densities of states.

## IVC. Gray-tin variants as 3D TIs

Gray Sn may be considered the parent of several families of TIs discussed in this section that occur in the zinc-blende type structure, and to which gray Sn is connected adiabatically. Among the group IVA elements, gray Sn with an inversion-symmetric diamond structure possesses a non-trivial band topology,(Fu and Kane 2007) while the lighter elements Si and Ge with the same structure are trivial insulators. Topology of bands in Si, Ge and Sn is controlled by states near $E_F$ at $\Gamma$. The $j=1/2$ $s$-like doublet lies near $E_F$ at $\Gamma$. The $j=1/2$ $s$-like doublet lies above the $p$-like fourfold degenerate $j=3/2$ levels in Si/Ge, but in gray Sn this natural order is inverted through a single band inversion at $\Gamma$, Figs. 12(a,b), yielding a nontrivial zero-gap semiconductor or semimetal. A TI phase in Sn can be realized by lifting the degeneracy of the $j=3/2$ states via a lattice distortion, Fig. 12(c). Notably, the band gap in the III-V zinc-blendes generally decreases for heavier constituent atoms with larger lattice constants, suggesting the presence of an inverted band gap. For example, first-principles computations predict TlP and TlAs to be stable in the zinc-blende structure with inverted bands at $\Gamma$. Similarly, the band gap of the trivial insulator InSb becomes non-trivial when the lattice is expanded sufficiently. (Lin, Das, Wang, et al. 2013; Ciftci, Colakoglu, and Deligoz 2008)

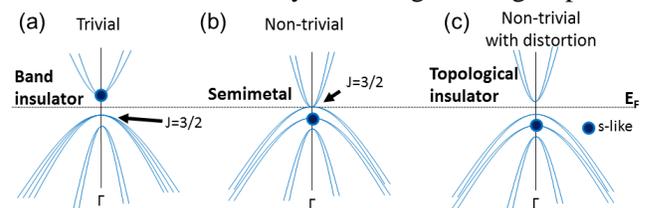

Fig. 12: Schematic band structures at Γ in Si/Ge (a), band inversion in Sn (b), and a TI realized by a distortion (c).

### IVC.1 Ternary half-Heusler compounds

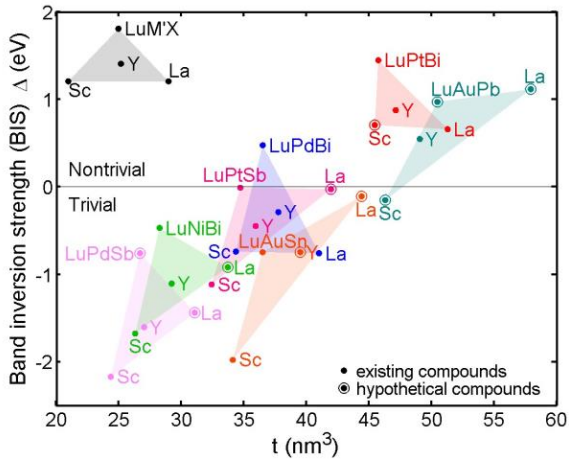

Fig. 13: Band inversion strength, $\Delta=\Gamma_8-\Gamma_6$, for various half-Heuslers versus $t=(Z_{M'}+Z_X)V$, where $Z_{M'}(Z_X)$ is atomic number of $M'(X)$ atom and $V$ is the unit-cell volume.(Lin, Wray, et al. 2010; Al-Sawai et al. 2010)

Chemically, the ternary half-Heuslers, $MM'X$, for $M = $ (Lu, La, Sc, Y) and $M'X = $ (PtBi, AuPb, PdBi, PtSb, AuSn, NiBi, PdSb), involve a total of 18 valence electrons per formula unit obtained by combining ten $d$ orbitals of $M'$ atom with two $s$- and six $p$-orbitals of the $X$ atom.(Chadov et al. 2010; Lin, Wray, et al. 2010; Xiao, Yao, et al. 2010; Al-Sawai et al. 2010) These 18 electrons can be accommodated in closed $d^{10}s^2p^6$ shells with zero total spin and angular momentum, and can thus, in principle, support a non-magnetic insulating band gap.

Because $M'$ and $X$ atoms $((M'X)^-)$ form a zinc blende type sublattice, the half-Heuslers resemble 3D-HgTe and InSb. As discussed in Sec. IIC, the half-Heuslers with inverted band structure at $\Gamma$ are connected adiabatically with nontrivial HgTe. The band topology in half-Heuslers is determined by the relative energies of the s-like $\Gamma_6$ and p-like $\Gamma_8$ levels, and the energy difference, $\Delta= E_{\Gamma_8} - E_{\Gamma_6}$, can be considered a measure of the band inversion strength (BIS). Fig. 13 shows the relationship between $\Delta$ and $t = (Z_{M'}+Z_X)V$, where $V$ is the cell volume, and $Z_{M'}(Z_X)$ is the atomic number of $M'(X)$ atom. $t$ captures effects on $\Delta$ of changes in the $(M'X)$ unit as well as the overall cell volume. Effects of disorder or non-stoichiometry could be modeled using a variety of approaches. (Bansil et al. 1981; Lin et al. 2006; Khanna et al. 1985; Huisman et al. 1981) The $\Delta=0$ line separates trivial and nontrivial phases. Materials near the zero line will be amenable to switching between trivial and non-trivial states with external perturbations.

Insight is gained by framing the four compounds containing the same binary $(M'X)$ unit with a triangle, Fig. 13.(Lin, Wray et al. 2010; Al-Sawai et al. 2010) Remarkably, for all seven subgroups, Sc, La, and Lu lie at corners of the triangle while Y lies inside. The 'orientation' of all seven triangles is the same in that it runs counterclockwise from La to Lu to Sc, where the element with the largest atomic mass, Lu, occupies the corner with the largest $\Delta$, except for the MAuSn and MAuPb subgroups. Moreover, the volume of compounds in each subgroup is ordered as Sc >Lu>Y>La, except for the MAuPb subgroup. The center of gravity for each triangle is seen to increase with $t$ for all subgroups. All compounds, whether physically realized or artificial, follow the aforementioned trends independent of the sign of $\Delta$. These relationships between $\Delta$ and $t$ may thus be valid more generally and useful in ascertaining the nature of topological phase in other nonmagnetic half-Heuslers. $\Delta$ would constitute a viable metric for genomic searches in this class of materials. YPtBi is a candidate for a topological superconductor.(Butch et al. 2011)

### IVC.2 Li₂AgSb class semiconductors

Like the half-Heuslers, $M'$ and $X$ atoms in the ternary intermetallics (Lin, Das, Wang, et al. 2013) Li₂$M'X$ [$M' = $ Cu, Ag, Au, or Cd; $X = $ Sb, Bi, or Sn] form a zinc-blende sublattice with a total of 18 valence electrons in closed $d^{10}s^2p^6$ shells. The electronic structure is similar to the half-Heuslers or gray Sn, and the band topology is controlled by the ordering of $\Gamma_6$ and $\Gamma_8$ levels. Band calculations predict Li₂AgBi and Li₂AuBi to be nontrivial semimetals, and $Li_2AgSb$ to be close to a critical point. The TI phase can be stabilized by a rhombohedral distortion with expansion along the hexagonal a-b plane, which lifts the degeneracy of $j = 3/2$ states and also induces a band inversion. $Li_2AsSb$ based compounds are connected adiabatically to nontrivial gray Sn. The phase diagram of Fig. 14 identifies $Li_2CdSn$ and InSb with expanded lattice as possible new nontrivial candidate TIs, demonstrating the value of band inversion strength and adiabatic continuity arguments as useful materials discovery tools.

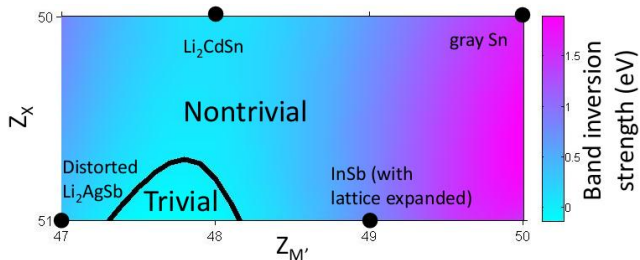

Fig. 14. Band inversion strength and topological phase diagram of $Li_2M'X$ as a function of atomic numbers of M' and X atoms.(Lin, Das, Wang, et al. 2013)

### IVC.3 Ternary Chalcopyrites, Famatinites, and Quaternary Chalcogenides

Ternary I-III-$VI_2$ and II-IV-$V_2$ chalcopyrites, (Feng et al. 2011) $I_3$–V–$VI_4$ famatinites, and quaternary $I_2$–II–IV–$VI_4$ chalcogenides (Wang, Lin, Das, et al. 2011) can be regarded as superlattices of distorted zinc-blende structure. Famatinites obey the octet rule and form a (I–$VI)_3$(V–VI) superlattice. Famatinites evolve into quaternary chalcogenides when one of their group-I elements is replaced by group-II and group-V elements by group-IV. The structure can be viewed as (I–$II)_2$(II–VI)(IV–VI) sublattice with two zinc-blende formula units. Substitution with larger atoms expands the lattice and increases both the tetragonal distortion and the crystal-field splitting. In contrast to the cubic zinc-blende type compounds in Sec. IVC above, the materials in this section naturally acquire a tetragonal distortion along the $c$-axis ($c < 2a$) through strong interlayer coupling between the two cation planes. This also results in a mismatch between the cation–anion bond lengths in two zinc-blende formula units, and helps lower the total energy. Tetragonal compression along the $c$-axis lifts the degeneracy of $p$-states in the zinc-blende lattice at $\Gamma$, and a subsequent inversion between the s- and p-states as in Fig.12(c) yields a TI or semimetal phase. (Feng et al. 2011)

### IVC.4 LiAuSe Honeycomb Lattice

Like the ternary cubic semiconductors, topological phases can be expected among the closed-shell relatives of graphene with graphite type XYZ structure for 8 or 18 valence electrons. Suitable combinations are (i) X = Li, Na, K, Rb or Cs, Y = Zn, Cd or Hg, and Z = P, As, Sb or Bi; (ii) X = K, Rb or Cs, Y = Ag or Au, and Z = Se or Te; (iii) X = rare earth, Y = Ni, Pd or Pt, and Z = P, As, Sb or Bi. Among these numerous possibilities, Zhang, Chadov, et al. (2011) consider LiAuSe, LiAuTe, CsAuTe, KHgBi, and CsHgBi due to the likelihood of being synthesized. Electronic structures are similar to their cubic counterparts except that in the binary semiconductors or $C_{1b}$ Heuslers, the bonds within the YZ tetrahedrons are of $sp^3$ or $sd^3$ type, while in the planar graphite-type layers the σ-type bonding occurs between the $sp^2$ or $sd^2$ orbitals. The remaining $p$ orbitals provide π-bonding similar to graphite. LiAuSe and KHgSb are both semimetals, although KHgSb is topologically trivial while LiAuSe is nontrivial.

### IVC.5 β-$Ag_2Te$

Ag-based chalcogenide $Ag_2Te$ undergoes a transition below 417 K from the α- to the β-phase, which is a narrow gap nonmagnetic semiconductor (Dalven 1966; Dalven and Gill 1967; Junod et al. 1977) with an unusually large and nonsaturating quasilinear magnetoresistance. (Xu et al. 1997) Such a large magnetoresistance cannot be explained within a conventional quadratic band structure, and a gapless linear dispersion driven by disorder effects was proposed. (Abrikosov 1998) First-principles calculations, however, predict β-$Ag_2Te$ to be a TI. (Zhang, Yu, et al. 2011) The high temperature α-phase of $Ag_2Te$, on the average, possesses an inversion-symmetric antifluorite structure with three interpenetrating fcc sublattices of Te, Ag(1) and Ag(2) with Te and Ag(1) forming a zinc-blende sublattice. Band structure of α-$Ag_2Te$ is similar to that of HgTe with an inverted band ordering at $\Gamma$. In the β-phase, $Ag_2Te$ assumes a distorted antifluorite structure in which the structural distortion removes the degeneracy at $\Gamma$ and opens a nontrivial insulating gap with the Dirac cone lying inside the gap. The emergence of metallic surface states is confirmed by the experimental observation of pronounced Aharonov-Bohm oscillations and a weak Altshuler-Aronov-Spivak effect in electron transport measurements on β-$Ag_2Te$ nanoribbons. (Sulaev et al. 2013) Pressure dependence of topological phase transitions in $Ag_2Te$ at room temperature is discussed by Zhao, Wang et al. (2013).

### IVD 2D Topological Materials

Conducting electrons at the edges of a 2D TI with single-Dirac-cone edge states can only move parallel or anti-parallel to the edges with opposite spins. In a 3D TI, on the other hand, although the surface states are free from backscattering, they can still scatter at other angles. 2D TIs with single Dirac-cone edge states are thus more promising for spintronics applications since the only scattering channel (backscattering) is prohibited. In the 2D TIs realized experimentally so far [HgTe/CdTe, InAs/GaSb quantum wells],(König et al. 2007; Knez, Du, and Sullivan 2011) the bandgap is very small so that transport measurements below 10K are required to see topological states. The need for finding 2D TIs with larger band gaps is thus clear, and DFT calculations are playing a major role in predicting possible new 2D TI materials.

To date, most predicted 2D TIs have been obtained by reducing dimensionality in quantum wells or in slabs of 3D TIs. In particular, 2D TIs are predicted in thin films of almost every class of 3D TIs [e.g., Bi/Sb, Sn, $Bi_2Se_3$, $Bi_2Te_3$, $Ge(Bi,Sb)_2Te_4$, Tl-based chalcogenides].(Liu et al. 2010; Wada et al. 2011; Chuang et al. 2013; Singh et al. 2013; Lin, Markiewicz et al. 2010) The topological characteristics are sensitive to thickness, composition and strain, and are tunable by electrical gating. Substrates can modify the electronic structure and change the band topology. For example, Bi thin films are predicted to be 2D TIs, but Sb thin films are trivial and become nontrivial under applied strain.(Wada et al. 2011; Chuang et al. 2013) Some of these thin films have been synthesized in various experiments but there still is no transport evidence for their being QSH insulators.(Zhang et al. 2010; Chun-Lei et al. 2013; Kim, Jin, et al. 2014)

A number of strategies for engineering topological states in 2D systems or their heterostructures have been proposed. Examples are: (i) Adding adatoms of heavy elements in the graphene structure to induce a stronger SOC for driving a topological phase transition (Weeks et al. 2011; Kane and Mele 2005a); (ii) Applying circularly polarized laser field on a 2D electronic system where the light field acts like a Rashba-type SOC to generate a non-trivial gap in an optical lattice (Inoue and Tanaka 2010); (iii) Stacking two Rashba-type spin-orbit coupled 2D electron gases with opposite signs of Rashba coupling in adjacent layers in heterostructure geometry (Das and Balatsky 2013); and, (iv) GaAs/Ge/GaAs heterostructure with opposite semiconductor interfaces acting as Rashba-bilayers to allow a band inversion yielding a 15 meV or larger insulating gap. (Zhang, Lou, et al. 2013)

**IVD.1 III-V HgTe/CdTe quantum well structures**

HgTe/CdTe quantum well is the first materials realization of a 2D topological phase. (Bernevig, Hughes, and Zhang 2006; König et al. 2007) HgTe and CdTe both possess the zinc-blende structure. HgTe is a topologically non-trivial semimetal with a single band inversion at $\Gamma$, Fig 12(b), while CdTe is topologically trivial, Fig. 12(a). However, the HgTe/CdTe heterostructure exhibits a thickness-dependent QSH state (Kane and Mele 2006) as first predicted theoretically, (Bernevig, Hughes, and Zhang 2006) and then verified experimentally. (König et al. 2007) This system has been reviewed extensively, (Hasan and Kane 2010; Qi and Zhang 2011) see the materials inventory in the Appendix for recent literature on HgTe/CdTe and InAs/GaSb/AlSb quantum well systems.

**IVD.2 Graphene**

Graphene is the natural starting point for discussing 2D systems considered in this section. Properties and applications of graphene have been reviewed extensively in the literature, (Beenakker 2008; Castro Neto et al. 2009; Peres 2010; Goerbig 2011; Das Sarma et al. 2011; Kotov et al. 2012) and therefore, we make only a few remarks here. Graphene is the simplest 2D topological system in which the theoretically predicted SOC gap is too small to be accessible experimentally. If the SOC is turned off, graphene becomes gapless with Dirac cones centered at the BZ corners K and K', which are not TRIM points. SOC only opens a band gap, but does not induce changes in the parity of eigenstates or band inversions at the TRIM points $\Gamma$ and M. Notably, any 2D electron gas can be transformed into a host for Dirac fermions via patterning with a periodic array of gates. (Kotov et al. 2012) In particular, a 2D electron gas under an external hexagonal periodic potential or optical lattice has been shown to develop graphene-like massless Dirac cones both theoretically (Park and Louie 2009; Wunsch, Guinea, and Sols 2008) and experimentally. (Gomes et al. 2012; Shikin et al. 2014) Introduction of adatoms on a graphene sublattice is another viable route for inducing a topological gap. (Hu et al. 2012)

**IVD.3 'Beyond' graphene atomically thin films: silicene, germanene, stanene**

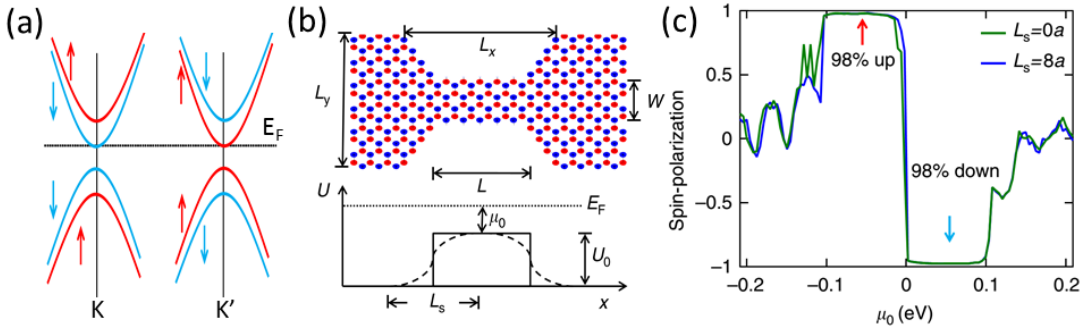

Fig. 15: (a) Schematic band structure of silicene in an external electric field. (b) Design of a spin-filter as a quantum point contact with barrier potential U(x). (c) Spin-polarization of the filter as a function of the barrier height μ₀. (Tsai et al. 2013)

Like a single layer of C atoms (graphene), Si, Ge and Sn can also form atomically thin crystals yielding silicene, germanene and stanene. Freestanding silicene and its Ge and Sn counterparts are 'beyond' graphene 2D materials with advantages over graphene of a stronger SOC and atomic bonds that are buckled and not flat like graphene. (Mas-Ballesté et al. 2011; Butler et al. 2013) First-principles calculations show these materials to be QSH insulators. Aside from being gapped, their band structures are similar to graphene with conduction and valence band edges located at the K and K' corners of the BZ. The SOC induced gaps in germanene and stanene are predicted to be 25meV and 70meV, respectively, large enough for room-temperature applications. (Liu, Feng, and Yao 2011; Tsai et al. 2013) The band gap can be tuned with a perpendicular electric field, which breaks the inversion symmetry due to the buckling of the honeycomb structure, (Drummond, Zólyomi, and Fal'ko 2012; Tsai et al. 2013) and topological phase transitions can be realized by gating control. The resulting field-tunable nondegenerate states possess nearly 100% spin-polarization, providing a basis for designing high efficiency spin filters (see Fig. 15), and other devices for manipulating spin currents. (Tsai et al. 2013; Gupta et al. 2014; Chuang et al. 2013) Although silicene has been grown on various substrates, the resulting electronic and geometric structures may be quite different from those of freestanding silicene films.(Vogt et al. 2012; Feng, Ding, et al. 2012; Fleurence et al. 2012; Meng et al. 2013) Honeycomb III-V thin films are a natural extension of silicene, and low-buckled GaBi, InBi, and TlBi thin films are predicted to be 2D TIs.(Chuang et al. 2014)

First-principles calculations predict films of Sn compounds SnX (X= H, I, Br, Cl, F or OH) to be in the QSH phase, (Xu, Yan, et al. 2013) with the hydrogenated version called stanane. Hybridization with the X atom removes $p_z$ bands from the $E_F$, so that low-energy states are dominated by the s and p orbitals, and the band topology is controlled by band inversion at Γ as in 3D gray Sn. An insulating gap as large as 300 meV is predicted in SnI. Similar results are found in films of other compounds. GeI also shows a gap of 300 meV. PbX (X=H, F, Cl, Br, I) is predicted to be a 2D TI, but phonon calculations suggest that the structure is not stable.(Si et al. 2014) A study of band topology and stability of the multilayer stanane indicates that one to three layer thick films are TIs, and films with greater thickness are metallic.(Chou et al. 2014) Functional thin films of Bi and Sb have also been predicted to be large gap TIs with large band gaps ranging from 0.74-1.08eV.(Song et al. 2014)

## IVE Organometallic compounds

Computations indicate the presence of TI phases in 2D-triphenyl-lead Pb(C₆H₅)₃ (Wang, Liu, and Liu 2013a) and 2D-organometallics such as π-conjugated Ni-bis-dithiolene Ni₃C₁₂S₁₂.(Wang, Su, and Liu 2013) Triphenyl-Pb has a buckled hexagonal structure in which the para-Pb atoms are displaced alternately up and down out of the plane of the benzene rings. Without SOC, the band structure displays a gapless Dirac cone at the K-point, but a gap of ~8.6 meV opens up when SOC is turned on. When Pb is replaced by magnetic Mn this family can exhibit QAH effect. (Wang, Liu, and Liu 2013b) In contrast to triphenyl-Pb, 2D Ni₃C₁₂S₁₂ adopts a kagome lattice. The band structure now contains two Dirac bands near $E_F$ with a gap of $\Delta_1 = 13.6$ meV, and a flat band, which is a distinctive feature of the kagome lattice, lying $\Delta_2 = 5.8$ meV above the upper Dirac band. Both $\Delta_1$ and $\Delta_2$ vanish in the absence of SOC. Liu et al. (Liu, Wang, et al. 2013) predict 2D In-phynylene to have a quasi-flat Chern band near $E_F$ in its ferromagnetic phase.

## IVF Transition Metal Compounds

### IVF.1 Iridates

$5d$ orbitals of Ir in iridium oxides provide a playground for unfolding the interplay between electron correlation and SOC effects. In addition to the Ir-skutterudites, topological materials have been proposed in iridates with pyrochlore, perovskite, and layered honeycomb structures. Ir-pyrochlores, $R_2Ir_2O_7$ (R = Nd, Sm, Eu, and Y), exhibit metal insulator transitions. (Matsuhira et al. 2007) LDA+U calculations predict the magnetic phase of $Y_2Ir_2O_7$ to be a Weyl semimetal. (Wan et al. 2011) Phase diagram of Ir-pyrochlores with varying strengths of electron-electron interaction and SOC harbors a variety of topological phases. (Pesin and Balents 2010) Crystal field splittings induced by lattice distortion of the $IrO_6$ octahedra can lead to topological insulator (Kargarian, Wen, and Fiete 2011) and topological Mott insulator (Yang and Kim 2010) states in pyrochlore iridates. In the honeycomb lattice of $Na_2IrO_3$, a QSH insulator phase and a fractionalized QSH state have been proposed. (Shitade et al. 2009; Young, Lee, and Kallin 2008) Ir-perovskites $Sr_{n+1}Ir_nO_{3n+1}$ undergo a transition from a Mott insulator (Kim et al. 2008) to a correlated metal with increasing $n$, while orthorhombic perovskite $SrIrO_3$ is metallic. Due to strong SOC, band structures of Ir-perovskites display narrow $J_{eff}=1/2$ bands near $E_F$. $SrIrO_3$ with a staggered potential in alternating layers could be a strong TI (Carter et al. 2012). The presence of topological phases in the iridates has not yet been confirmed experimentally. (Li et al. 2013; Ye et al. 2013)

**IVF.2 Osmium Compounds**

Os-compounds involve 5d electrons like the iridates, but the SOC is stronger in Os compared to Ir. Os-compounds are thus natural candidates to search for correlated topological phases. Os-oxides $AOs_2O_4$, where $A$ is an alkali atom such as Mg, Ca, Sr, or Ba, have a metastable spinel structure which is predicted by DFT + U calculations to host a plethora of exotic phases depending on the strength of $U$.(Wan, Vishwanath, and Savrasov 2012) In particular, ferromagnetic $CaOs_2O_4$ and $SrOs_2O_4$ are predicted to be magnetic axion insulators with $\theta=\pi$, protected by inversion symmetry with a gapped surface phase. Yan, Muchler *et al.* (2012) propose that Ce-filled skutterudite compounds $CeOs_4As_{12}$ and $CeOs_4Sb_{12}$ are zero gap TIs with an inverted band gap between Os-$d$ and Ce-$f$ orbitals. When Kondo effect is included between these orbitals, the systems could yield a topological Kondo insulator.

**IVG Heavy f-electron Materials**

**IVG.1 Topological Kondo insulator (TKI) $SmB_6$**

$SmB_6$ is a mixed valence compound whose insulating gap arises from $f$-$d$ hybridization. (Martin and Allen 1979; Martin and Allen 1981) Optical (Gorshunov et al. 1999) and transport (Flachbart et al. 2001) studies give evidence of two gaps of ~3-5 meV and ~10-20 meV. The transition to the insulating state starts below 50K, but the conductivity remains finite and saturates below 4K suggesting the presence of metallic states within the gap. (Allen, Batlogg, and Wachter 1979; Cooley et al. 1995) These observations led to renewed interest in $SmB_6$ as a possible TKI system. (Dzero et al. 2010; Dzero et al. 2012; Kim, Xia, and Fisk 2014) Other predicted TKI's include SmS, (Li, Li, et al. 2014; Zhao, Lu, et al. 2014) and $YB_6$ and $YB_{12}$ as topological Kondo crystalline insulators.(Weng, Zhao, et al. 2014)

ARPES experiments on $SmB_6$ show a spin-polarized surface state at $\Gamma$ lying inside the bulk band gap (Denlinger et al. 2013) and X-centered electron-like bands spanning the gap, (Xu, Shi, et al. 2013; Neupane, Alidoust, et al. 2013; Jiang, Li, et al. 2013) including earlier hints of in-gap states. (Miyazaki et al. 2012) Ionic character of bonding in $SmB_6$ implies the possibility of surface terminations with opposite polarity, which could drive surface reconstructions, a variety of which are seen in STM measurements.(Yee et al. 2013; Rößler et al. 2013) Other explanations for the observed in-gap states in $SmB_6$ suggest these to be of bulk origin (Frantzeskakis et al. 2013) or as being polarity driven non-topological states. (Zhu, Nicolaou, et al. 2013) Transport results such as the observed linear magnetoresistance (Thomas et al. 2013) and quantum Hall effect (Kim, Thomas, et al. 2013) in $SmB_6$ have been interpreted within a TKI framework.

**IVG.2 Topological Mott insulators in actinides**

In going from $4f$ electrons in Sm to $5f$ electrons in U, Pu or Am, Kondo physics becomes too weak to support an insulating phase,(Das, Zhu, and Graf 2012; Das, Durakiewicz, et al. 2012) but a Mott insulator can be induced via strong Coulomb interaction. LDA+U calculations (Zhang, Zhang, Wang, et al. 2012) predict AmN and PuY, which crystallize in the rocksalt structure, to be TIs where the low energy electronic spectrum is governed by the actinide $f$ and $d$ orbitals. The 6d states split into $\Gamma_7^+$ and $\Gamma_8^+$ and 5f into $5f^{5/2}$ and $5f^{7/2}$ due to effects of SOC and crystal field. At a certain value of the Coulomb interaction $U = 2.5$ eV, the $5f^{7/2}$ states shift away from $E_F$ and create a non-trivial insulator with band inversion along the $\Gamma$-X direction.

LDA+DMFT calculations predict $PuB_6$, which crystallizes in the same CsCl-type structure as $SmB_6$, to be a topological Mott insulator. (Deng, Haule, and Kotliar 2013) The band orderings are similar to PuY or AmN with a surface Dirac cone located in the inverted band gap at the X-point. We should keep in mind that the predicted size of the insulating gap depends on the value of $U$ used in the computations. A quantum phase transition in the spin-orbit channel is predicted in $URu_2Si_2$ yielding a topological metal. (Das 2012) An experimental confirmation of a topological Mott insulator in the actinides has not been reported.

## IVH Weyl and 3D Dirac Semimetals

Weyl and 3D Dirac semimetal phases have been predicted in a variety of existing or tailored systems as follows: (i) Pyrochlore iridates could host 24 Weyl nodes through the interplay of SOC and strong electron correlations (Wan et al. 2011); (ii) Ferromagnetic $HgCr_2Se_4$ spinel with nodes protected by $C_4$ point group symmetry (Xu, Weng, et al. 2011; Fang et al. 2012); (iii) Crystalline $Cd_3As_2$ (Wang, Weng, et al. 2013) and $A_3Bi$ (A = Na, K, Rb) (Wang, Sun, et al. 2012); (iv) β-cristobalite $BiO_2$;(Young et al. 2012) (v) strained $Hg_{1-x-y}Cd_xMn_yTe$ under magnetic field (Bulmash, Liu, and Qi 2014); (vi) $TlBiSe_2$ (Singh et al. 2012); (vii) Heterostructure of magnetically doped 3D TI and normal insulator (Burkov and Balents 2011b); and, (viii) A superlattice of alternating layers with odd and even parity orbitals. (Das 2013) Experimentally realized 3D Dirac semimetals so far are $Cd_3As_2$ (Neupane, Xu, et al. 2014; Borisenko et al. 2014; Ali et al. 2014) and $Na_3Bi$ (Liu, Zhou, et al. 2014; Xu, Liu, et al. 2015). Bulk Dirac cones have also been reported in $SrMnBi_2$ and $CaMnBi_2$ (Feng et al. 2013). Attempts to obtain Weyl fermions by breaking TRS in 3D Dirac semimentals have been undertaken in $Cd_3As_2$.(Jeon et al. 2014) By breaking inversion symmetry rather than the TRS, Weyl semimetals have been predicted and then realized in TaAs family.(Huang et al. 2015; Weng et al. 2015; Xu, Belopolski, et al. 2015; Lv et al. 2015; Xu, Alidoust, et al. 2015; Yang, Liu, et al. 2015; Zhang et al. 2015)

## IVI Other topological materials

### IVI.1 Complex oxides

Complex oxides are widely studied in the context of magnetism, metal-insulator transitions, superconductors, (Pickett 1989; Imada, Fujimori, and Tokura 1998; Gardner, Gingras, and Greedan 2010) and as platforms for oxide electronics. (Tokura and Hwang 2008; Heber 2009) First-principles calculations indicate that the electronic structures of bulk $YBiO_3$,(Jin et al. 2013) $BaBiO_3$ (Yan, Jansen, and Felser 2013), (111) bilayer of $LuAlO_3$ (Xiao et al. 2011), and a superlattice of

$CrO_2/TiO_2$ (Cai et al. 2015) are potential TI and QAH candidates. [Iridates are related materials discussed in Sec. IVF.1 above.] The band inversion between Bi- and O-$p$ states in $YBiO_3$ upon the inclusion of SOC occurs at TRIM points R, rather than at Γ. (Jin et al. 2013) $BaBiO_3$ is a known superconductor with $T_c \sim 30$ K, (Sleight, Gillson, and Bierstedt 1975; Cava et al. 1988) which is predicted by DFT to have a large topological band gap of 0.7 eV. (Yan, Jansen, and Felser 2013) Although the treatment of electron correlation effects is a source of uncertainty in band structure computations, Okamoto *et al*. (2014) show that the TI phase survives in DMFT computations on a $LuAlO_3$ bilayer.

### IVH.2 Skutterudites, antiperovskites, other structures

We note a variety of other theoretically predicted topological materials as follows: (i) $IrBi_3$(Yang and Liu 2014), and $CoSb_3$ (Smith et al. 2011) in their skutterudite crystal structure are predicted to have nontrivial topological phase by DFT+U calculations. (ii) Antiperovskite compounds ($M_3N$)Bi ($M$=Ca,Sr, and Ba) (Sun et al. 2010) and $Ca_3PbO$ family (Kariyado and Ogata 2011); (iii) Cubic perovskites $CsXI_3$ ($X$=Pb, Sn) under strain (Yang et al. 2012); (iv) Strain induced topological phase transition in Zintl compounds $Sr_2X$ ($X$=Pb, Sn) (Sun et al. 2011); (v) Single-layer $ZrTe_5$ and $HfTe_5$ in layered structure as 2D TIs with band gap as large as 0.4 eV (Weng, Dai, and Fang 2014); (vi) β-GaS and GaS-II family under uniaxial strain (Huang, Zhang, et al. 2013); and (vii) $Cs(Sn,Pb,Ge)(Cl,Br,I)_3$ ternary halides (Yang et al. 2012).

## V. Outlook and Conclusions

Although considerable progress has been made in discovering new topological materials during the last few years, the menu of choices available currently is still quite limited, especially when it comes to 2D systems and to materials where topological states are protected through combinations of time-reversal, crystalline and particle-hole symmetries. Implementation of robust tools for assessing topological characteristics of band structures (e.g., $Z_2$ invariants, Chern numbers, Berry connection, Berry curvature) in widely available open source band structure codes will help accelerate the discovery process, and the development of viable materials with flexibility and tunability necessary for fundamental science investigations and as platforms for various applications. Limitations of the band theory in strongly correlated materials apply to the topological band theory as well, but extensions of the weak coupling band theory to reliably treat the intermediate coupling regime with predictive capabilities should be possible. Realistic modeling of the spectral intensities, including matrix element effects, is needed for reliable identification of the key topological states and their spin textures, especially for the surface sensitive angle-resolved photoemission and scanning tunneling spectroscopies, which are the most relevant spectroscopies in connection with topological materials. In this vein, realistic modeling of transport properties of topological materials, including effects of external fields, electron-phonon couplings, defects and impurities is another area that needs attention for developing practical spintronics and other applications. There are many experimental challenges as well in synthesizing materials, which can reach the topological transport regime. A related challenge is realizing high-degree of theoretically predicted spin-polarization of the topological surface states, which is reduced by various intrinsic and extrinsic effects. There are exciting possibilities for observing in a solid state setting non-abelian particles associated usually with high-energy physics such as axions, Majorana fermions, magnetic monopoles, and fractional excitations. The challenge here is to control correlated physics of magnetism and superconductivity within the topological matrix. Nevertheless, there can be little doubt that vast treasures of riches and surprises await us in the goldmine of 3D and 2D topological materials, and their interfaces and heterostructures involving magnetic, non-magnetic and superconducting materials.

**Acknowledgments**

It is great pleasure to acknowledge our collaborations and discussions on various aspects of topological materials with the following colleagues: J. Adell, N. Alidoust, J. M. Allred, T. Balasubramanian, B. Barbiellini, L. Balicas, S. Basak, A. Balatsky, I. Belopolski, G. Bian, M. Bissen, J. Braun, R. S. Cava, H. R. Chang, T.-R. Chang, Y. Chen, F. C. Chou, J. G. Checkelsky, Y.-T. Cui, F. C. Chuang, J.-W. Deng, J. D. Denlinger, C. Dhital, J. H. Dil, H. Ding, K. Dolui, W. Duan, T. Durakiewicz, H. Ebert, A. V. Fedorov, Z. Fisk, L. Fu, D. R. Gardner, Q. Gibson, M. J. Graf, D. Grauer, G. Gupta, M. Z. Hasan, Y. He, J. Hoffman, Y. S. Hor, D. Hsieh, T. H. Hsieh, C.-H. Hsu, C.-Yi Huang, Y.-B. Huang, Z.-Q.. Huang, E. Hudson, Z. Hussain, A. Taleb-Ibrahimi, Y. Ishida, M. B. Jalil, H.-T. Jeng, H. Ji, S. Jia, Y. Jo, S. Kaprzyk, S. Khadka, D.-J. Kim, T. Kondo, J. W. Krizan, S. K. Kushwaha, G. Landolt, M. Leandersson, Y. S. Lee, C. Liu, G. C. Liang, J. Liu, Y.-T. Liu , Z. Liu, M. Lindroos, J. Lee, V. Madhavan, R. S. Markiewicz, D. Marchenko, A. Marcinkova, P. E. Mijnarends, J. Minár, K. Miyamoto, S.-K. Mo, R. Moore, E. Morosan, F. Meier, M. Neupane, J. Nieminen, Y. Ohtsubo, Y. Okada, T. Okuda, N. P. Ong, J. Osterwalder, A. Pal, L. Patthey, A. Petersen, R. Prasad, C. M. Polley, D. Qian, O. Rader, A. Richardella, M. Serbyn, B. Singh, F. von Rohr, R. Shankar,  B. Slomski, N. Samarth, J. Sánchez-Barriga, M. R. Scholz, M. Severson, F. Schmitt, A. Sharma, S. Shin, B. Singh, Z. X. Shen, A. Soumyanarayanan, W.-F. Tsai, A. Varykhalov, A. Volykhov, D. Walkup, Y.-J. Wang, Z. Wang, S. D. Wilson, L. A. Wray, D. Wu,Y. Xia, J. Xiong, S. Xu, H. Yan, L. V. Yashina, M. M. Yee, D. Zhang, Y. Zhang, Y. Wang, Z. Wang, Bo Zhou, W. Zhou. The work of AB was supported by the U. S. Department of Energy, Basic Energy Sciences, Division of Materials Sciences grant numbers DE-FG02-07ER46352 (core research), DE-AC02-05CH11231 (computational support at NERSC) and DE-SC0012575 (work on layered materials). Work of HL and TD was supported by award number NRF-NRFF2013-03 of the Singapore National Research Foundation.

# Appendix: Inventory of 2D and 3D Topological Materials

| Material | *References (Th/Exp denote theory/experiment focus) |
|---|---|
| **Bi/Sb variants** | |
| Sb;Bi$_{1-x}$Sb$_x$ | Th: Fu 2007, Teo 2008, Zhu 2014, Sahin 2015; <br> Exp: Hsieh 2008, Roushan 2009, Zhu 2013, Nishide 2010, Guo 2011, Soumyanarayanan 2013, Hsieh 2009, Gomes 2009, Tian 2015 |
| As | Th: Campi 2012 |
| Bi$_2$Se$_3$;Bi$_2$Te$_3$;Sb$_2$Te$_3$ | Th: Zhang 2009, Xia 2009, Luo 2012, Koleini 2013, Li, 2014, Wu 2013b, Chis 2012, Hsieh 2009, Chen 2009, Wang 2013, Wang 2010, Kou 2013, Hinsche 2012, Bahramy 2012, Menshchikova 2011, Wu 2013a, Zhang 2012, Luo 2013, Eremeev 2013, Henk 2012, Henk 2012, Niesner 2012, Wan 2014, Hinsche 2015, Wang 2015; <br> Exp: Xia 2009, Hsieh 2009, Chen 2009, Hsieh 2009, Hor 2009, Aguilar 2012, Analytis 2010, Kirshenbaum 2013, Jozwiak 2013, Luo 2013, Crepaldi 2012, Aitani 2013, Pauly 2012, Zhu 2013, Kim 2013, Jenkins 2012, Fauqué 2013, Deshpande 2014, Nomura 2014, Luo 2013, Kong 2013, Kim 2012, Yan 2013, Zhu 2013, Crepaldi 2012, Qu 2010, Zhang 2012, Wang 2013, Rischau 2013, Tian 2013, Ning 2013, Qu 2012, Okada 2012, Zhu 2011, Chiu 2013, Zhang 2012, Wang 2012, Zhu 2011, Kuroda 2010, Kim 2011, Zhang 2010, Alpichshev 2012b, Zhang 2009, Jenkins 2010, Chen 2010, Steinberg 2010, Hoefer 2014, Mellnik 2014, Li 2014, Zhao 2014, Liu 2014, Fu 2014, Bansal 2014, Kim 2014, Hong 2014, Boschker 2014, Lang 2014, Sung 2014, Yan 2014, Sessi 2014, Zhao 2014, Neupane 2014, Kastl 2015, Park 2015, Kim 2014, Vargas 2014, Edmonds 2014, Tsipas 2014, Cacho 2015, Seibel 2015 |
| Bi$_2$Te$_2$Se; B$_2$X$_2$X'(B = Bi, Sb; X,X' = S, Se, Te) | Th: Lin 2011, Wang 2011, Chang 2011, Menshchikova 2011, Gehring 2013, Eremeev 2012; <br> Exp: Ren 2010, Xiong 2012, Wang 2014, Ren 2012, Bao 2012, Miyamoto 2012, Barreto 2014, Hajlaoui 2014 |
| Alloy (Bi,Sb)$_2$(S,Se,Te)$_3$ | Th: Niu 2012; <br> Exp: Zhang 2011, Kong 2011, Taskin 2011, Ji 2012, Shikin 2014, Yoshimi 2014, Lee 2014, Tang 2014, Ando 2014, Ou 2014, Yoshimi 2015, Yang 2015 |
| Mn/Fe/Cr/Gd/V magnetically doped | Th: Zhang 2012; <br> Exp: Hor 2010, Checkelsky 2012, Xu 2012, Zhang 2012, Beidenkopf 2011, Jiang 2013, Chang 2013, Zhao 2013, Kou 2013, Harrison 2014, Okada 2011, Schlenk 2012, Chen 2010, Wray 2011, Zhang 2013, Lee 2015, Fan 2014, Chang 2015, Checkelsky 2014, Sessi 2014, Bestwick 2015, Li 2015 |
| Cu doped | Exp: Kriener 2012, Wray 2010, Lawson 2012 |
| (Bi$_{1-x}$In$_x$)$_2$Se$_3$ | Exp: Wu 2013, Brahlek 2012 |
| Ge$_m$Bi$_{2n}$Te$_{m+3n}$ ; A$_m$B$_{2n}$X$_{m+3n}$ (A = Pb, Sn, Ge; B = Bi, Sb; X,X' = S, Se, Te) | Th: Xu 2010, Jin 2011, Sa 2011, Kim 2010, Sa 2014, Singh 2013a; <br> Exp: Xu 2010, Okuda 2013, Marcinkova 2013, Neupane 2012, Taskin 2011, Nomura 2014, Muff 2013, Okamoto 2012, Niesner 2014 |
| (PbSe)$_5$(Bi$_2$Se$_3$)$_{3m}$ | Exp: Nakayama 2012, Fang 2013, Sasaki 2014 |
| TlBiSe$_2$; MM'X$_2$ (M = Tl, M' = Bi or Sb, and X = Te, Se, or S) | Th: Lin 2010, Yan 2010, Chang 2011, Singh 2012, Niu 2012, Eremeev 2010, Eremeev 2011; <br> Exp: Sato 2010, Chen 2010b, Xu 2011, Sato 2011, Kuroda 2010, Shoman 2015, Xu 2015, Novak 2015 |
| Bi$_2$TeI | Th: Klintenberg 2010, Tang 2014 |
| BiTeX (X=I, Br, Cl) | Th: Bahramy 2012, Landolt 2013, Eremeev 2012, Bahramy 2011; <br> Exp: Landolt 2013, Chen 2013, Ishizaka 2011, Crepaldi 2014, Tran 2014 |
| LaBiTe$_3$ | Th: Yan 2010 |
| (Bi$_2$)$_m$(Bi$_2$Te$_3$)$_n$ | Th: Jeffries 2011;<br>Exp: Jeffries 2011, Valla 2012, Shirasawa 2013 |
| Bi$_{14}$Rh$_3$I$_9$ | Th: Rasche 2013;<br>Exp: Pauly 2015 |
| | |
| **Topological crystalline insulator (TCI)** | |
| TCI: SnTe;(Pb,Sn)(S,Se,Te) | Th: Hsieh 2012, Wang 2013, Safaei 2013, Wang 2014, Fang 2014, Wang 2014, Sun 2013, Hota 2013, Eremeev 2014, Liu 2014, Tang 2014, Wrasse 2014;<br>Exp: Xu 2012, Dziawa 2012, Tanaka 2012, Okada 2013, Littlewood 2010, Takafuji 1982, Iizumi 1975, Burke 1965, Li 2013, Balakrishnan 2013, Taskin 2014, Liang 2013, Polley 2014, Zhong 2013, Zeljkovic 2015, Zeljkovic 2014, Shen 2014, Zeljkovic 2015 |
| **Sn variants** | |
| Sn | Th: Fu 2007, Küfner 2013;<br>Exp: Barfuss 2013 |
| 3D HgTe | Th: Chiu 2012, Beugeling 2015, Ortix 2014;<br>Exp: Orlita 2014, Oostinga 2013, Olshanetsky 2015, Sochnikov 2015, Ren 2014 |
| Half-Heusler | Th: Chadov 2010, Lin 2010, Xiao 2010, Al-Sawai 2010;<br>Exp: Miyawaki 2012, Wang 2013 |
| Li$_2$AgSb; Li$_2$M'X [M= Cu, Ag, Au, or Cd and X = Sb, Bi, or Sn] | Th: Lin 2013 |
| Ternary Chalcopyrites; Famatinites; and Quaternary Chalcogenides I-III-VI2 ,II-IV-V2, I3–V–VI4,I2–II–IV–VI4,I-III-VI2 and II-IV-V2 | Th: Feng 2011, Wang 2011 |
| LiAuSe; KHgSb | Th: Zhang 2011 |
| β-Ag$_2$Te | Th: Zhang 2011;<br>Exp: Sulaev 2012, Zhao 2013 |
| **2D systems** | |
| Graphene | Th: Kane 2005a, Beenakker 2008, Peres 2010, Lau 2013, Ghaemi 2012, Wunsch 2008, Kotov 2012, Goerbig 2011, Sarma 2011, Weeks 2011, Diniz 2013, Maher 2013, Kou 2013,  Hu 2013, Park 2009, Pesin 2012, Neto 2009, Vaezi 2013, Hu 2012, Chang 2014;<br>Exp: Beenakker 2008, Peres 2010, Kotov 2012, Goerbig 2011, Sarma 2011, Maher 2013, Neto 2009, Gomes 2012, Kravets 2013, Ju 2015, Young 2014,  Gorbachev 2014 |
| Silicene;germanene | Th: Liu 2011, Zhang 2013, Tabert 2013, Kikutake 2013, Drummond 2012, Tsai 2013, Liu 2011, Ezawa 2013, Ezawa 2012, Tahir 2013, Gupta 2014, Padilha 2013,  Pan 2014 |
| GaAs/Ge/GaAs | Th: Zhang 2013 |
| GaBi; InBi; TlBi | Th: Chuang 2014, Li 2015 |
| Bi4Br4 | Th: Zhou 2014 |
| GeX;SnX;PbX;BiX(X= H, I, Br, Cl, F or OH) | Th: Si 2014, Zhou 2014 |
| Bi | Th: Murakami 2006, Wada 2011, Liu 2011, Huang 2013, Wang 2014, Ma 2015;<br>Exp: Jnawali 2012, Hirayama 2011, Hirahara 2012, Wang 2013, Wells 2009, Lükermann 2012, Coelho 2013, Chun-Lei 2013, Kim 2014, Drozdov 2014, Lu 2015, Takayama 2015 |
| Sb | Th: Zhang 2012, Chuang 2013;<br>Exp: Yao 2013, Bian 2011 |
| BiTeX (X=I, Br, Cl) | Th: Kou 2014 |
| HgTe/CdTe | Th: Bernevig 2006, Luo 2010, Khaymovich 2013;<br>Exp: König 2007, Brüne 2012, König 2013, Nowack 2013, Zholudev 2012, Gusev 2014, Hart 2014 |
| InAs/GaSb/AlSb | Th: Liu 2008;<br>Exp: Knez 2011, Knez 2014, Du 2015 |
| $LaAlO_3/SrTiO_3$;$SrTiO_3/SrIrO_3$;<br>$KTaO_3/KPtO_3$ | Th: Lado 2013;<br>Exp: Cheng 2013 |
| $LaAuO_3$; $SrIrO_3$ | Th: Xiao 2011, Okamoto 2013 |
| $CrO_3/TiO_3$ | Th: Cai 2013 |
| Transition metal dichalcogenides | Th: Qian 2014, Cazalilla 2014 |
| **Organometallic** | Th: Liu 2013, Wang 2013a, Wang 2013b, Wang 2013, Li 2014 |
| **Transition metal compounds** | |
| $R_2Ir_2O_7$ (R = Nd, Sm, Eu, and Y) | Th: Wan 2011, Pesin 2010, Yang 2010, Kargarian 2011;<br>Exp: Matsuhira 2007, Ueda 2014 |
| $Pb_2Ir_2O_{7-x}$ | Th&Exp: Hirata 2013 |
| $Na_2IrO_3$; $Li_2IrO_3$ | Th: Young 2008, Shitade 2009, You 2012, Kargarian 2012;<br>Exp: Alpichshev 2015 |
| $Sr_{n+1}Ir_nO_{3n+1}$;$Sr_2IrO_3$;$Sr_2IrRhO_6$ | Th: Carter 2012;<br>Exp: Li 2013, Kim 2008, Ye 2013 |
| $AOs_2O_4$, (A=Mg, Ca, Sr) | Th: Wan 2012 |
| $CeOs_4As_{12}$ ; $CeOs_4Sb_{12}$ | Th: Yan 2012 |
| $ZrTe_5$; $HfTe_5$ | Th: Weng 2014 |
| **Heavy f-electron Materials** | |
| $SmB_6$, $YB_6$ and $YB_{12}$ | Th: Dzero 2012, Dzero 2010, Lu 2013, Zhu 2013, Weng 2014;<br>Exp: Zhu 2013, Frantzeskakis 2013, Xu 2013, Flachbart 2001, Rößler 2013, Yee 2013, Miyazaki 2012, Jiang 2013, Denlinger 2013, Neupane 2013, Kim 2013, Kim 2014, Zhang 2013, Thomas 2013, Li 2014, Syers 2015, Fuhrman 2015, Neupane 2015, |
| SmS | Th: Li 2014, Zhao 2014 |
| PuB6 | Th: Deng 2013 |
| AmN and PuY, | Th: Zhang2012 |
| $URu_2Si_2$ | Th: Das 2012 |
| **Weyl and 3D Dirac Semimetals** | |
| $HgCr_2Se_4$ | Th: Fang 2012, Xu 2011 |
| $A_3Bi$ (A = Na, K, Rb) | Th: Wang 2012, Narayan 2014, Gorbar 2015;<br>Exp: Liu 2014, Xu 2015 |
| $BiO_2$ | Th: Young 2012 |
| $Hg_{1-x-y}Cd_xMn_yTe$ | Th: Bulmash 2014 |
| $Cd_3As_2$ | Th: Wang 2013a;<br>Exp: Borisenko 2013, Neupane 2013, Ali 2014, Liu 2014, Jeon 2014, Pariari 2015, Yi 2014 |
| $SrMnBi_2$; $CaMnBi_2$ | Exp: Wang 2011, Feng 2013 |
| LaAgSb2 | Exp: Wang 2012 |
| TaAs, TaP, NbAs, NbP | Th: Huang 2015; Weng 2015<br>Exp: Xu 2015, Lv 2015, Xu 2015, Yang 2015, Zhang 2015 |
| **Complex oxides** | |

| YBiO₃ | Th: Jin 2013 |
|---|---|
| BaBiO₃ | Th: Yan 2013 |
| **Skutterudites** | |
| IrBi₃ | Th: Yang 2013 |
| CoSb₃ | Th: Smith 2011 |
| **Antiperovskites** | |
| M₃NBi (M=Ca,Sr,Ba) | Th: Sun 2010 |
| Ca₃PbO | Th: Kariyado 2011 |

\* In the interest of brevity, only the last name of the first author is given. However, in cases of multiple references with the same name and year, titles of the articles should allow the reader to access the appropriate literature. Although we have attempted to separate references as being theoretical or experimental in their focus, many articles involve both components. This inventory should not be considered exhaustive, although it should be fairly complete as of the submission date, and includes some subsequent updating.